\documentclass[times,number,5p]{elsarticle}
\pdfoutput=1

\usepackage{graphicx,miller,bm,booktabs,float,nicefrac,nicematrix,url,color,microtype,subcaption,lipsum,multirow}
\usepackage[export]{adjustbox}

\usepackage{siunitx}
\usepackage[version=4]{mhchem}
\usepackage[breaklinks,hidelinks]{hyperref}
\usepackage[ruled]{algorithm2e}
\usepackage[T1]{fontenc}

\biboptions{numbers,sort&compress}

\DeclareMathAccent{\wtilde}{\mathord}{largesymbols}{"65}
\newcommand*{\matr}[1]{\bm{\mathit{#1}}}

\DeclareSIUnit{\pixel}{pixels}
\DeclareSIUnit{\voxel}{voxels}
\DeclareSIUnit\weightpercent{\text{wt.}\%}

\DeclareUnicodeCharacter{03B1}{$\alpha$}
\DeclareUnicodeCharacter{03B3}{$\gamma$}
\DeclareUnicodeCharacter{025B}{$\varepsilon$}
\DeclareUnicodeCharacter{03B5}{$\varepsilon$}

\journal{Acta Materialia}

\begin{document}

\begin{frontmatter}

\title{Grain-level effects on in-situ deformation-induced phase transformations in a complex-phase steel using 3DXRD and EBSD}

\author[a,b]{James A. D. Ball}
\author[c]{Claire Davis}
\author[c]{Carl Slater}
\author[a]{Himanshu Vashishtha}
\author[a]{Mohammed Said}
\author[d]{Louis Hébrard}
\author[d]{Florian Steinhilber}
\author[e]{Jonathan P. Wright}
\author[b]{Thomas Connolley}
\author[b]{Stefan Michalik}
\author[a]{David M. Collins} \ead{D.M.Collins@bham.ac.uk}

\affiliation[a]{organization={School of Metallurgy and Materials, University of Birmingham},
        addressline={Edgbaston}, 
        city={Birmingham},
        postcode={B15~2TT}, 
        country={United~Kingdom}}
\affiliation[b]{organization={Diamond Light Source Ltd.},
        addressline={Harwell Science and Innovation Campus}, 
        city={Didcot},
        postcode={OX11~0DE}, 
        country={United~Kingdom}}
\affiliation[c]{organization={WMG, University of Warwick}, 
        city={Coventry},
        postcode={CV4~7AL}, 
        country={United~Kingdom}}
\affiliation[d]{organization={METAL group, INSA Lyon},
        addressline={25 Avenue Jean Capelle}, 
        city={69621 Villeurbanne},
        country={France}}
\affiliation[e]{organization={European Synchrotron Radiation Facility (ESRF)},
        addressline={71 Avenue des Martyrs},
        city={38000 Grenoble},
        country={France}}

\begin{abstract}
A novel complex-phase steel alloy is conceived with a deliberately unstable austenite, $\gamma$, phase that enables the deformation-induced martensitic transformations (DIMT) to be explored at low levels of plastic strain.
The DIMT was thus explored, in-situ and non-destructively, using both far-field Three-Dimensional X-Ray Diffraction (3DXRD) and Electron Back-Scatter Diffraction (EBSD).
Substantial $\alpha'$ martensite formation was observed under \qty{10}{\percent} applied strain with EBSD, and many $\varepsilon$ grain formation events were captured with 3DXRD, indicative of the indirect transformation of martensite via the reaction $\gamma \rightarrow \varepsilon \rightarrow \alpha'$.
Using $\varepsilon$ grain formation as a direct measurement of $\gamma$ grain stability, the influence of several microstructural properties, such as grain size, orientation and neighbourhood configuration, on $\gamma$ stability have been identified.
Larger $\gamma$ grains were found to be less stable than smaller grains.
Any $\gamma$ grains oriented with \hkl{100} parallel to the loading direction preferentially transformed with lower stresses.
Parent $\varepsilon$-forming $\gamma$ grains possessed a neighbourhood with increased ferritic/martensitic volume fraction.
This finding shows, unambiguously, that $\alpha$/$\alpha'$ promotes $\varepsilon$ formation in neighbouring grains.
The minimum strain work criterion model for $\varepsilon$ variant prediction was also evaluated, which worked well for most grains.
However, $\varepsilon$-forming grains with a lower stress were less well predicted by the model, indicating crystal-level behaviour must be considered for accurate $\varepsilon$ formation.
The findings from this work are considered key for the future design of alloys where the deformation response can be controlled by tailoring microstructure and local or macroscopic crystal orientations.
\end{abstract}

\begin{keyword}
Martensitic phase transformation \sep Micromechanics \sep 3D characterization \sep Austenitic stainless steels
\end{keyword}

\end{frontmatter}

\section{Introduction}
Contemporary Advanced High-Strength Steel (AHSS) alloys, such as dual-phase (DP), complex-phase (CP), quenching and partitioning (Q\&P) and Transformation-Induced Plasticity (TRIP) steels, exploit a deformation-induced martensitic transformation (DIMT) to achieve both high strength and ductility \citep{kobayashi_energy_2009}.
In these steel alloys, a deformation-induced phase transformation from face-centered cubic retained austenite (FCC $\gamma$) to body-centered tetragonal martensite (BCT $\alpha'$) is combined with other deformation modes including dislocation slip and mechanical twinning to achieve substantial total plasticity, allowing the material to work-harden significantly and thereby increasing ductility \citep{bressanelli_effects_1966, zackay_enhancement_1967}. 
To achieve this response, judicious alloy design is employed to achieve a suitably unstable austenite for the TRIP effect to initiate during loading, whilst remaining sufficiently stable for the DIMT to occur over a large plastic strain range \citep{herrera_design_2011}.
Given its role in controlling the macroscopic mechanical properties of the alloy; a mature understanding of what affects austenite grain stability is key to new alloy development \citep{jimenez-melero_effect_2009}.

When TRIP steels are subjected to applied stresses above the global yield, retained austenite can transform to $\alpha'$ martensite via one of two possible routes.
The $\alpha'$ martensite can form directly from $\gamma$ \citep{dash_martensite_1963, olson_mechanism_1972, hedstrom_stepwise_2007} at grain boundaries and triple points \citep{das_morphologies_2008, tian_comparing_2018, he_mechanisms_2022}.
An indirect two-stage transformation is also possible, governed by deformation bands formation along $\hkl{111}$ planes in the $\gamma$ grain.
This is typical of TRIP alloys with a low stacking fault energy (SFE); here the \hkl{111}\hkl<110> perfect dislocations will dissociate into Shockley partials of type \hkl{111}\hkl<112> with a stacking fault between them.
The formation of \hkl{111} type deformation bands are created due to the successive formation of these faults as plasticity builds \citep{lecroisey_martensitic_1972}.
In low SFE alloys, these faults preferentially form on every second $\hkl{111}$ plane, the regions of which are now crystallographically distinct, well known as the intermediate hexagonal close-packed (HCP) martensite phase, $\varepsilon$ martensite \citep{venables_martensite_1962, olson_general_1976-2}.
The $\alpha'$ phase can then form from the intersection of these shear bands \citep{bogers_partial_1964, olson_mechanism_1972, olson_kinetics_1975, olson_general_1976-2}, within individual shear bands \citep{staudhammer_nucleation_1983, tian_deformation_2017, tian_comparing_2018}, or at the intersection of shear bands and grain boundaries \citep{mangonon_martensite_1970, murr_effects_1982, staudhammer_nucleation_1983}.
It is widely accepted that the kinetically favoured formation route follows the transformation $\gamma \rightarrow \varepsilon \rightarrow \alpha'$ \citep{mangonon_martensite_1970}.
Recent work by \citeauthor{tian_deformation_2017} has demonstrated that a relationship exists between the underlying stability of the austenite grain against transformation and the specific transformation mechanism observed \citep{tian_deformation_2017, tian_comparing_2018}, making it possible to probe austenite grain stability by identifying the martensite transformation pathway.

Austenite grain stability is known to depend on many material and microstructural properties, such as alloy chemistry \citep{otte_formation_1957, cina_transitional_1958}, stacking fault energy \citep{venables_martensite_1962, staudhammer_nucleation_1983, talonen_effect_2007, tian_comparing_2018, molnar_stacking_2019}, grain size \citep{rigsbee_inhibition_1979, jimenez-melero_martensitic_2007, sohrabi_deformation-induced_2020, kang_situ_2023}, grain orientation \citep{magee_transformation_1966, hedstrom_elastic_2008, das_estimation_2011, neding_formation_2021}, and grain neighbourhood \citep{hedstrom_deformation_2005, hedstrom_stepwise_2007, jimenez-melero_situ_2011, li_situ_2014, tomida_variant_2021, wang_ebsd_2021, toda_multimodal_2022, kang_situ_2023}.
However, the exact correlations between austenite grain stability and some of these properties are still under dispute.
Recent work by \citet{turteltaub_grain_2006, haidemenopoulos_kinetics_2014, jung_effect_2011, jimenez-melero_martensitic_2007} have shown that larger austenite grains are less stable than smaller austenite grains, and that refining the overall retained austenite grain size has a beneficial effect on stability against transformation.
However, explorations of both very small (\qty{> 1}{\micro\metre}) \citep{somani_enhanced_2009, marechal_linkage_2011, kisko_influence_2013} and very large grains (\qtyrange[range-phrase=--,range-units=single]{52}{284}{\micro\metre}) \citep{shrinivas_deformation-induced_1995} disagree with these findings and show that larger grains can be more stable against transformation.

Measuring individual austenite grain stability during in-situ deformation is challenging and requires sophisticated experimental design and data analysis.
As such, in-situ studies of austenite grain stability are limited -- recent notable work has leveraged in-situ electron back-scatter diffraction (EBSD) \citep{kim_strain_2022, zhang_microstructure_2022, kang_situ_2023}, neutron diffraction \citep{das_influence_2018, onuki_situ_2020, polatidis_high_2020} and synchrotron X-ray diffraction (SXRD) \citep{hedstrom_deformation_2005, hedstrom_stepwise_2007, li_situ_2014, tian_martensite_2017, tian_micromechanics_2018, neding_formation_2021}.
In-situ explorations of austenite grain stability at the individual grain level are limited further still, and usually consider only a handful of grains, primarily due to methodological and technological limitations \citep{hedstrom_deformation_2005, hedstrom_stepwise_2007, hedstrom_elastic_2008, li_situ_2014, neding_formation_2021, zhang_microstructure_2022, kang_situ_2023}.

This study seeks to rectify the lack of mesoscale explorations of the DIMT by utilising two in-situ experimental techniques -- EBSD and Three-Dimensional X-ray Diffraction (3DXRD).
A novel multi-phase stainless steel alloy, recently characterised in a previous paper \citep{ball_registration_2023}, is developed to facilitate substantial deformation-induced martensitic transformations at low applied strains, enabling in-situ characterisation with far-field 3DXRD at the European Synchrotron Radiation Facility (ESRF).
An advanced 3DXRD reconstruction pipeline recently developed at Diamond Light Source (DLS) \citep{ball_implementing_2022, ball_per-grain_2023}, is employed to facilitate large-scale multi-phase analysis of 3DXRD data collected at multiple load steps.
Direct spatially resolved microstructure measurements were obtained from an in-situ EBSD experiment with tensile deformation, to further elucidate the transformation behaviour applied strains greater than those reachable by 3DXRD.

\section{Experimental Method}
\subsection{Material}
To investigate the DIMT, the 304L steel alloy was identified as a suitable candidate as it possesses a well established deformation-induced $\gamma \rightarrow \alpha'$ transformation.
At room temperature under uniaxial tension, the initiation strain for martensite formation is $>0.1$ \citep{hecker_effects_1982}; this represents a difficulty for 3DXRD which cannot accurately measure per-grain information at strain levels this high.
To address this shortcoming, a new steel was designed, bespoke for this study, that enables such a DIMT to be studied, but within the observable strain range permitted by 3DXRD.

A modified composition of 304L steel, hereon referred to as 304LM steel, was designed to destabilise the $\gamma$ phase, promoting the $\gamma \rightarrow \alpha'$ transformation at a lower macroscopic strain.
A matrix of compositions with varying Ni (5-10 wt.\%), Mn (0-4 wt.\%), Cr (19-21\%) and C (0.01-0.04 wt.\%) was simulated using the CALPHAD software Thermo-Calc 2021 using the FCFE7/MOBFE2 databases.
A down selection of suitable alloys was determined based on compositions where: (1) 100\% $\gamma$ is possible but with (2) the A$_{\textrm C3}$-A$_{\textrm C4}$ heat treatment window as small as is practicable, (3) the martensitic start temperature, $M_{\textrm s}$, is as high as possible but, critically, is below room temperature, (4) the transformation of $\gamma \rightarrow \alpha$ can be avoided during cooling, as predicted by Time-Temperature Transformation (TTT) diagrams, and (5) a stacking fault energy ($\gamma_{\rm SFE}$) that does not differ significantly from the unmodified 304L alloy composition.
Here, this is calculated via \citep{olson_general_1976-2}:

\begin{equation}
    \gamma_{\rm SFE} = n\rho \left( \Delta G^{\rm chem} + E^{\rm strain} \right) + 2\sigma(n)
\end{equation}

\noindent where $\Delta G^{\rm chem} = \Delta G^{\gamma \leftrightarrow \varepsilon}$, calculated by Thermo-Calc, and $\rho$ is the density of atoms along a \{111\} fault, calculated by \citet{wang_stacking_2020}:

\begin{equation}
    \rho = \frac{4}{\sqrt{3}}\frac{1}{a^2_{\rm FCC}N_A}
\end{equation}

\noindent For an austenitic stainless steel, $n = 2$, representing the number of atomic rows per stacking fault, the strain energy $E^{\rm strain} = 0$, the lattice parameter $a_{\rm FCC} = \qty{0.36}{\nano\metre}$ and the interfacial energy, $\sigma = \qty{8}{\milli\joule\per\metre\squared}$ \citep{curtze_thermodynamic_2011}.
$N_A$ is the Avogadro constant.

Following the aforementioned downselection criteria, the chosen composition is given in Table~\ref{tab:304l_modified_composition} \citep{ball_registration_2023}. The selected alloy was cast in a large billet followed by rolling, then solution annealed for \SI{12}{\hour} at \SI{1250}{\celsius} to increase the grain size, then quenched in air to room temperature.

\begin{table}[H]
\centering
\begin{tabular}{@{}l|llllllll@{}}
\textbf{Element} & C  & Ni & Cr & Mn & Si & P    & S    & Fe   \\ \midrule
\textbf{wt.\%}  & 0.04 & 7 & 19 & 2  & 1  & 0.04 & 0.03 & Bal.
\end{tabular}
\caption{Composition of alloy 304LM, selected using data obtained from Thermo-Calc thermodynamic equilibrium calculations. \citep{ball_registration_2023}.}
\label{tab:304l_modified_composition}
\end{table}

The as-received microstructure of this material has been extensively characterised with the use of Diffraction-Contrast Tomography (DCT) and ex-situ EBSD in a previous study \citep{ball_registration_2023}, the results of which will be briefly described here.
The final material had a multi-phase microstructure comprising \qtyrange[range-units=single,range-phrase=--]{\sim 60}{80}{\percent} retained austenite, with the remainder a mixture of polygonal ferrite and athermal martensite.
Grain size distributions gave a mean grain diameter of \qty{\sim 15}{\micro\metre}.
The retained austenite phase is broadly untextured, whilst the ferrite and martensite phases have a significant \hkl{111}\hkl<112> texture.
The combined polygonal ferrite and athermal martensite has a mean grain size of \qty{9}{\micro\metre}, calculated from EBSD measurements.

\subsection{EBSD with in-situ deformation}
In-situ EBSD samples were machined from the cast 304LM billet via wire-EDM to a tensile dogbone geometry.
Samples were polished in progressively finer grades of \ce{SiC} paper from \num{400} to \num{4000} grit, followed by a colloidal silica polish to \qty{0.04}{\micro\metre} surface finish, then a final electro-polish at \qty{20}{\celsius} with an 80:20 mixture of ethanol and perchloric acid, at a voltage of \qty{15}{\volt} for 20~seconds with a flow rate of \qty{20}{\litre\per\minute}.

\begin{figure}
    \centering
    \includegraphics[width=\columnwidth]{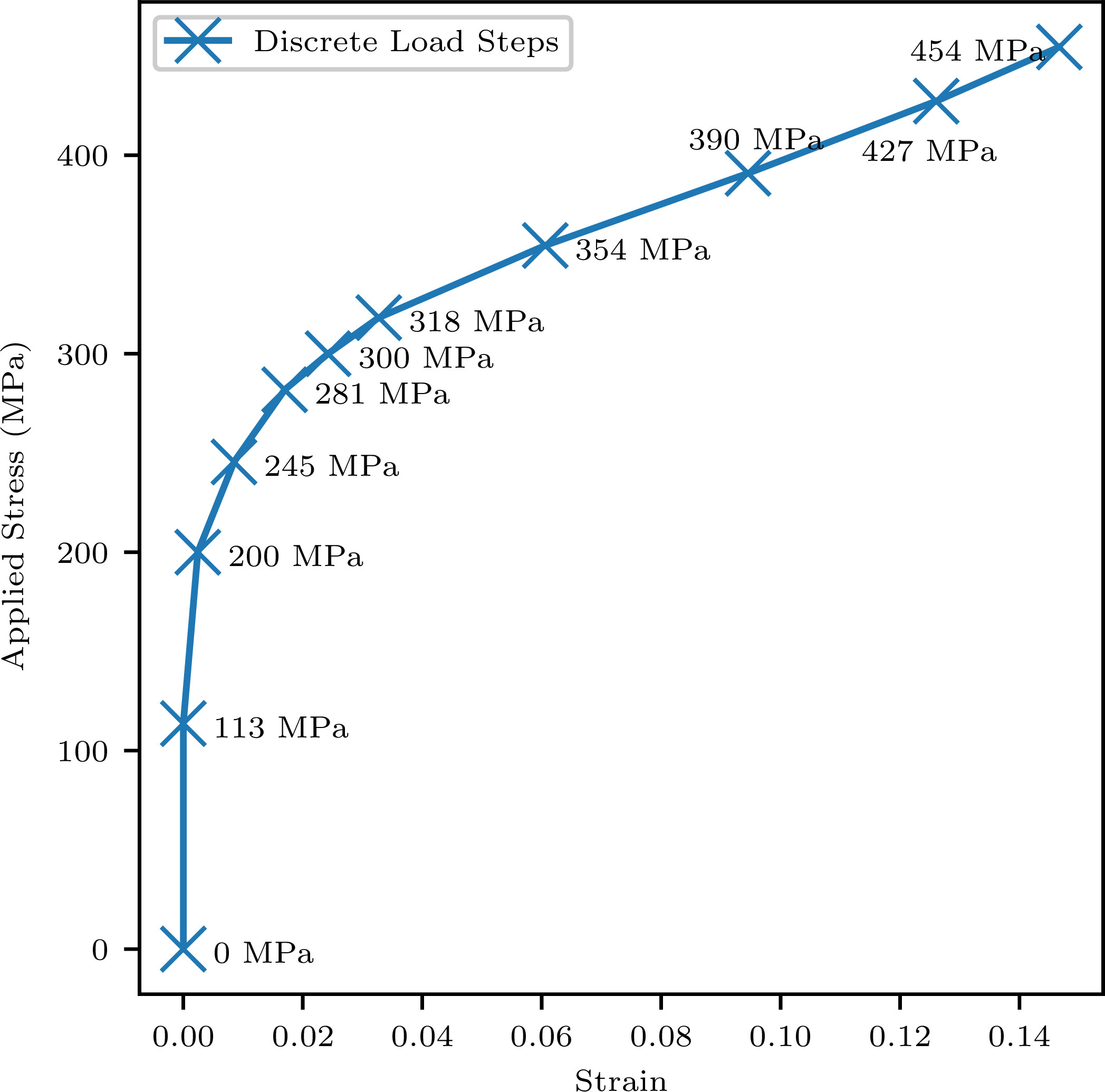}
    \caption{Macromechanical response of alloy 304LM with interrupted loading points during in-situ EBSD measurements.}
    \label{fig:stress_strain_ebsd}
\end{figure}

In-situ EBSD measurements were taken using a ThermoFischer Apreo 2 S HiVac FEG-SEM equipped with an Oxford Instruments Symmetry S3 EBSD detector. Interrupted-tensile loading at a series of stress targets (shown in Figure~\ref{fig:stress_strain_ebsd}) was applied using an in-situ Deben Microtest MT2000 tensile stage.
At each deformation step, the sample was re-aligned to the beam, and EBSD patterns were collected in a \qtyproduct[product-units=single]{250 x 250}{\micro\metre\squared} area with a \qty{1}{\micro\metre} step size, with data collection and pattern indexing conducted with the Oxford Instruments AZtec software. The diffraction patterns were acquired at a $1244 \times 1024$ pixel resolution, and were saved for offline data analysis.
For the final load step, a larger area EBSD map was collected without saving patterns over a \qtyproduct[product-units=single]{418.75 x 287}{\micro\metre\squared} area at a step size of \qty{0.25}{\micro\metre}. For all data acquired, the FEG-SEM was operating at \qty{20}{\kilo\electronvolt} with a \qty{6.4}{\nano\ampere} probe current.

\subsubsection{EBSD data analysis}
Analysis of EBSD data was performed with the MTEX MATLAB library \citep{bachmann_texture_2010}.
An initial grain segmentation was performed with a \qty{10}{\degree} misorientation tolerance.
Grains with an area of fewer than 3 pixels (\qty{3}{\micro\metre\squared}) were unassigned, then the segmentation was run again.
Kernel-average misorientation maps were calculated with a tolerance of \qty{2.5}{\degree}.
Geometrically necessary dislocation (GND) density maps were also calculated using the high-resolution EBSD (HR-EBSD) analysis software, CrossCourt \citep{wilkinson_mapping_2009}.
Here, the method utilises the measured lattice curvature to calculate GND density \citep{arsenlis_crystallographic_1999} by solving Nye’s dislocation tensor \citep{nye_geometrical_1953}.
Established methods exist to obtain GND density field estimations from EBSD measurements \citep{pantleon_resolving_2008, wilkinson_determination_2010}.
The reported GND density maps indicate the total dislocation density from all possible dislocation line vector combinations with a Burgers vector for the FCC austenite, $\gamma$, phase.
It is assumed that all dislocations are perfect.

\subsection{3DXRD with in-situ deformation}
\begin{figure}[ht!]
    \centering
    \includegraphics[width=\columnwidth]{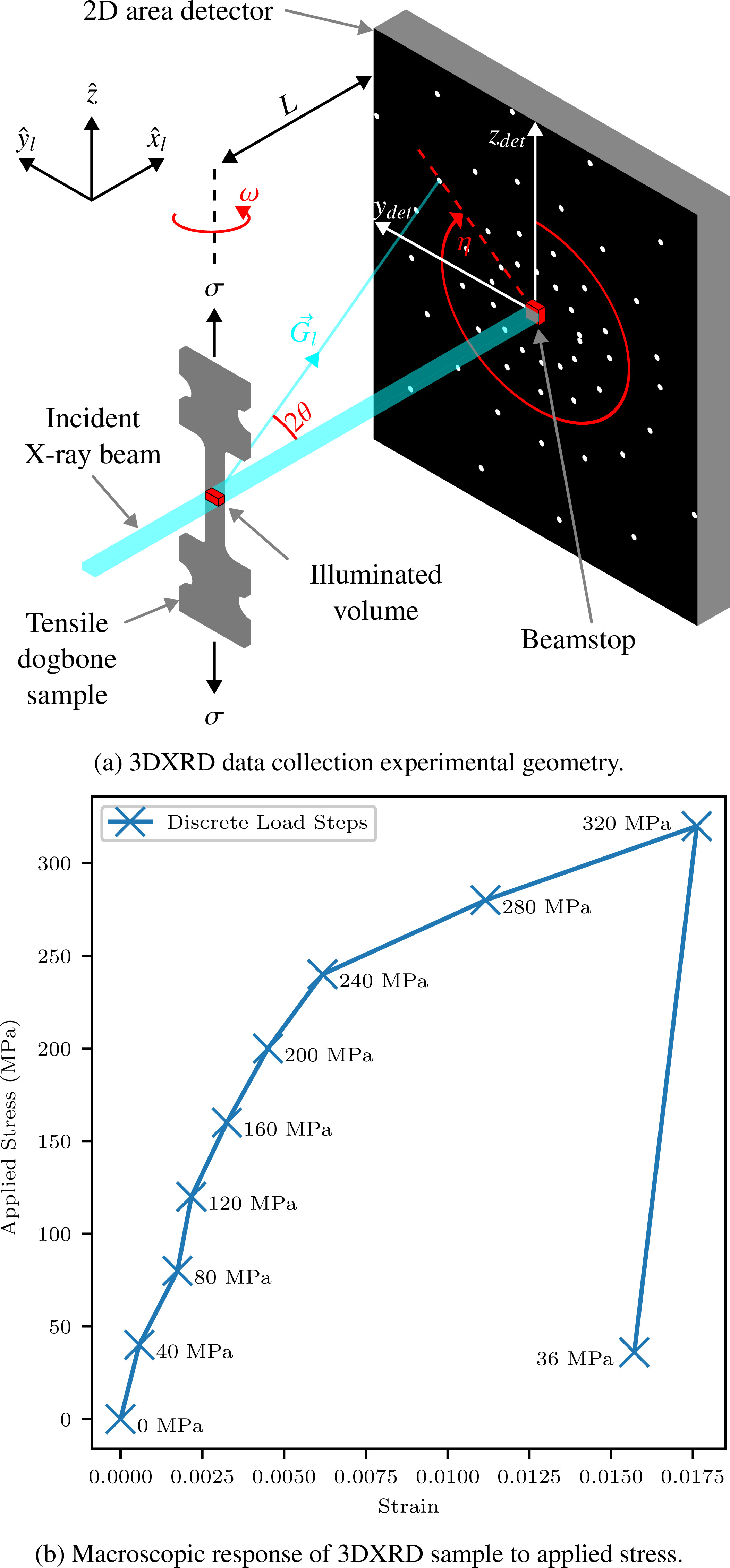}
    \caption{Geometry (a) and interrupted loading strategy (b) for 3DXRD measurements taken at the ID11 beamline, ESRF.}
    \label{fig:3dxrd_geom_and_stess_strain}
\end{figure}

In-situ 3DXRD samples were machined from the cast billet via wire-EDM to a tensile dogbone geometry with a \qtyproduct[product-units=single]{0.5 x 0.5}{\milli\metre\squared} gauge cross-section and a gauge length of \qty{2.39}{\milli\metre}.
Interrupted-loading far-field 3DXRD data were collected at the ID11 beamline, ESRF, as per Figure~\ref{fig:3dxrd_geom_and_stess_strain}a.
An incident monochromatic X-ray beam with a beam energy of \qty{44.0}{\kilo\electronvolt} and dimensions of \qty{1.0}{\milli\metre} (horizontal) by \qty{0.20}{\milli\metre} (vertical) illuminated the full width of the sample.
Diffracted X-ray beams were detected by a \numproduct{2048 x 2048}~px Frelon4M detector with a pixel size of \qty{50}{\micro\metre}, placed at a sample-detector distance, $L$, of \qty{145}{\milli\metre}.
Beam energy and sample-detector distance were calibrated with a \ce{CeO2} reference calibrant.
A Nanox tensile load frame \citep{gueninchault_nanox_2016} was used to apply a series of increasing tensile stresses on the sample, following a load step sequence outlined in Figure~\ref{fig:3dxrd_geom_and_stess_strain}b.
At each load step, four "letterbox" scans were collected, where the sample was rotated in a fly scan by \qty{360}{\degree} in $\omega$.
Detector images were collected every \qty{0.25}{\degree} in $\Delta\omega$.
After each "letterbox" scan, the sample was translated vertically by \qty{0.20}{\milli\metre}, creating a total illuminated volume of \qtyproduct[product-units=single]{0.5 x 0.5 x 0.8}{\milli\metre\cubed}.
After increasing the load for each load step, the sample was approximately re-aligned to the beam using a near-field camera.

\subsection{XRD Data Analysis}
\subsubsection{Quantitative phase assessment}
A Rietveld refinement using the software TOPAS \citep{coelho_topas_2018} was performed on the Frelon detector images recorded during each load step, following the procedure below:
\begin{enumerate}
\item Take the mean of every detector frame across every letterbox scan in the load step.
\item Perform an azimuthal integration of the result using the pyFAI Python package \citep{kieffer_new_2020} and export to a .xy file.
\item Import the.xy file into TOPAS.
\item Perform a Rietveld refinement with TOPAS following guidance by \citet{daniel_analysing_2023}.
\item Extract volume fractions from Rietveld result.
\end{enumerate}

\subsection{3DXRD Data Analysis}
The 3DXRD data analysis strategy followed procedures recently established at DLS \citep{ball_implementing_2022, ball_per-grain_2023}.
\subsubsection{Indexing pipeline}
\begin{figure}
    \centering
    \includegraphics[width=\columnwidth]{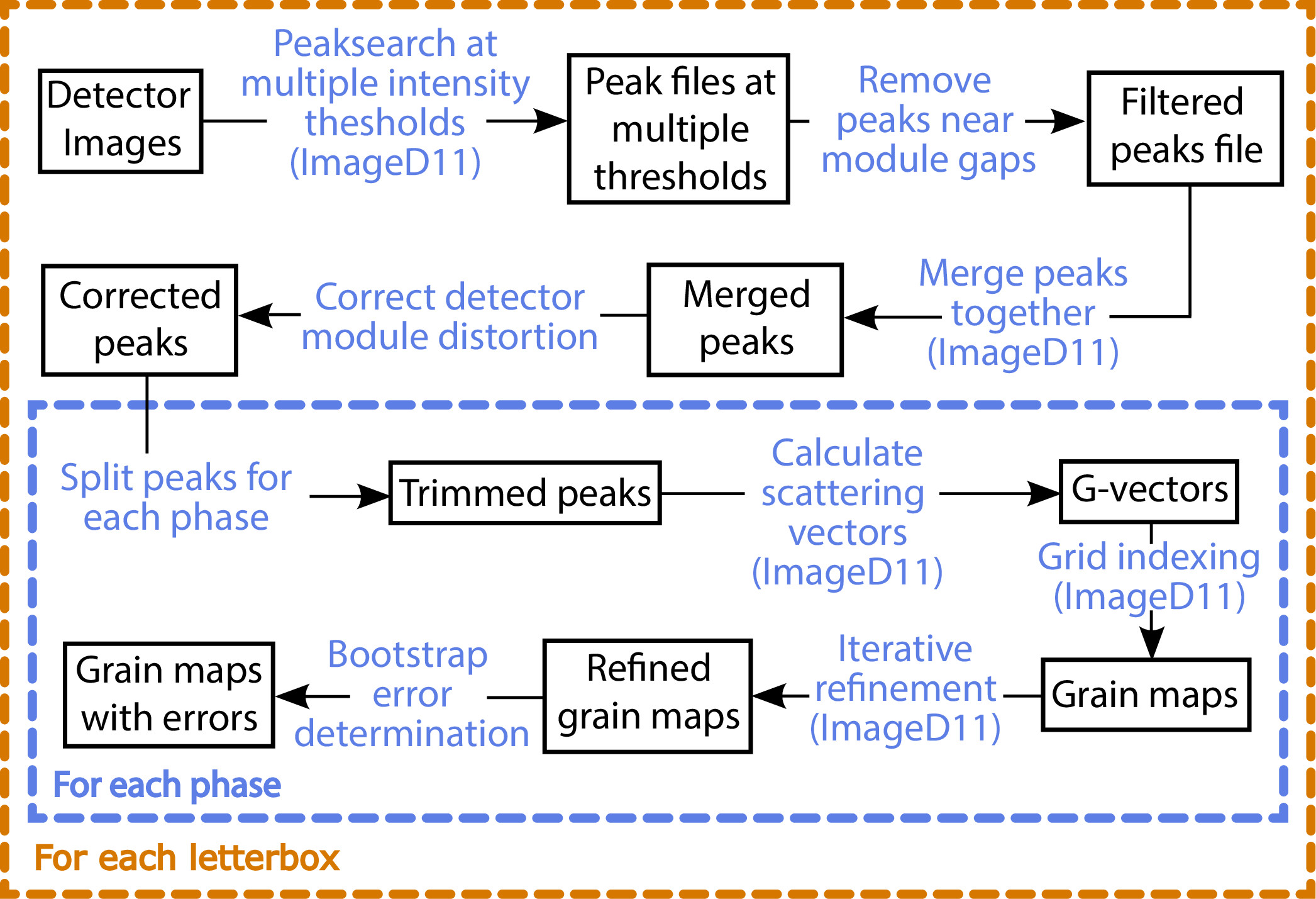}
    \caption{3DXRD data analysis procedure \citep{ball_per-grain_2023}.}
    \label{fig:3dxrd_analysis_process}
\end{figure}

An initial indexing and refinement pipeline, shown in Figure~\ref{fig:3dxrd_analysis_process}, was employed to extract individual grain positions, orientations, strain states and relative volumes, utilising the ImageD11 Python library \citep{wright_fable-3dxrdimaged11_2020} for peak searching, indexing and refining each letterbox scan.

Grain parameter errors were calculated using a "bootstrap" error detection routine, utilising multiple parameter refinements per grain with different sub-sets of diffraction peaks to probe the degree of convergence of the minimisation routine employed by ImageD11.
The reader is referred to the first implementation of this technique \citep{ball_per-grain_2023} for a more detailed description of the underlying error algorithm.

\subsubsection{Lattice parameters and stiffness constants}
The reference stress-free lattice parameters and stiffness constants were used as provided in Table~\ref{tab:lattice_constants}.
For $\varepsilon$, the HCP stiffness constants were calculated from the $\gamma$ stiffness constants using the method outlined by \citep{richeton_mechanical_2019}.

\begin{table}[h]
\centering
\begin{tabular}{@{}lccc@{}}
\toprule
\multirow{2}{*}{\textbf{Elastic constant}} & \multicolumn{3}{c}{\textbf{Phase}} \\
 & \textbf{$\gamma$} & \textbf{$\alpha$/$\alpha'$} & \textbf{$\varepsilon$} \\ \midrule
$C_{11}$ (\unit{\giga\pascal})& 204 & 231.4 & 266.8 \\
$C_{12}$ (\unit{\giga\pascal})& 133 & 134.7 & 130.5 \\
$C_{13}$ (\unit{\giga\pascal})&  &  & 72.7 \\
$C_{33}$ (\unit{\giga\pascal})&  &  & 324.7 \\
$C_{44}$ (\unit{\giga\pascal})& 126 & 116.4 & 65.7 \\ 
$a$ (\unit{\angstrom}) & 3.5925 & 2.872 & 2.541 \\
$c$ (\unit{\angstrom}) & 3.5925 & 2.872 & 4.140 \\ \bottomrule
\end{tabular}
\caption{Elastic and lattice constants used for $\gamma$ \citep{kluczynski_hot_2020}, $\alpha$/$\alpha'$ \citep{inal_second-order_2004}, $\varepsilon$ \citep{richeton_mechanical_2019} phases. $a$ and $c$ values for $\alpha'$ are assumed to be equal for 3DXRD indexing purposes.}
\label{tab:lattice_constants}
\end{table}

\subsubsection{Post-processing pipeline}
To stitch together the individual letterbox scans into contiguous volumes, and track grains across multiple load steps, a post-processing pipeline also developed at DLS was utilised.
Within each letterbox scan, duplicate grains were removed if their centre-of-mass positions and misorientation were within \qty{100}{\micro\metre} and \qty{1.0}{\degree} respectively, following the approach outlined by \citet{louca_accurate_2021} using an optimized form of the misorientation function found in the pymicro Python library \citep{proudhon_pymicro_2021}.
When stitching letterbox scans together, a modified form of the above approach was employed with separate distance tolerances in $xy$ (\qty{100}{\micro\metre}) vs $z$ (\qty{200}{\micro\metre}) and the same misorientation tolerance.
When tracking grains across multiple load steps, the original duplicate grain detection was used with  \qty{100}{\micro\metre} and \qty{1.0}{\degree} tolerances.

\subsubsection{Sample reference frame alignment}
Before tracking grains across load steps, consistent sample reference frames were created for each load step that removed residual sample-beam misalignments caused by sample movement while increasing the applied load.
Grain positions in the first load step were used as the ground truth reference.
At each load step, a temporary list of grains were generated that could be matched to reference grains using the duplicate grain detection algorithm previously described.
The sample rotation between load steps was determined using grain orientations, taking advantage of the high orientation-space precision afforded by far-field 3DXRD \citep{nervo_comparison_2014}.
An initial sample rotation was generated for each load step, and a minimisation routine was employed following the DIRECT algorithm \citep{jones_lipschitzian_1993, gablonsky_locally-biased_2001} implemented by the scipy Python package \citep{virtanen_scipy_2020}.
Here, the mean misorientation was minimised to obtain a ground truth value across all grain pairs by varying the sample rigid body rotation in Euler space.
With the sample rigid body rotation determined, updated grain positions were generated with the sample rotation taken into account.
The pycpd Python package \citep{gatti_pycpd_2022} was then employed to perform a rigid body transformation between the reference grain positions and the grain positions at the given load step.
This allowed the translation of the sample to be accurately determined between load steps.

\subsubsection{Macroscopic strain determination}
Due to the size of the tensile rig used for the 3DXRD experiments, it was not possible to determine macroscopic strain in-situ.
To determine macroscopic strain values for Figure~\ref{fig:3dxrd_geom_and_stess_strain}b, the centre-of-mass positions of grains within the illuminated volume were tracked across each load step, then an affine registration was performed using the pycpd Python library \citep{gatti_pycpd_2022} between tracked grain positions at the first load step and each subsequent load step.
The scale parameters were extracted from each \numproduct{4 x 4} affine transformation matrix by taking the length of the first three column vectors \citep{geometrian_given_2015}.
Given a \numproduct{4 x 4} affine transformation matrix:
\begin{equation}
\begin{bNiceMatrix}[columns-width=auto]
a & b & c & d\\
e & f & g & h\\
i & j & k & l\\
0 & 0 & 0 & 1
\end{bNiceMatrix}
\end{equation}
the column vector lengths were extracted:
\begin{equation}
\begin{aligned}
    s_x &= \left\Vert\left\langle a, e, i \right\rangle\right\Vert \\
    s_y &= \left\Vert\left\langle b, f, j \right\rangle\right\Vert \\
    s_z &= \left\Vert\left\langle c, g, k \right\rangle\right\Vert \\
    \vec{s} &= \left\langle s_x, s_y, s_z \right\rangle
\end{aligned}
\end{equation}
to give a scale vector $\vec{s}$.
The macroscopic strain tensor element $\epsilon_{zz}$ at each load step was then taken as $s_z-1$.

\subsubsection{Austenite to epsilon martensite transformation detection}
The list of all trackable austenite grains (that appeared in more than one load step) was used as the input to the detection algorithm.
For each tracked $\gamma$ grain $g_{\gamma}$, a list of its $\gamma$ grain representations $[g_{\gamma}^i]$ at each load step was generated.
For each $[g_{\gamma}^i]$ element at a given load step, a list of candidate $\varepsilon$ grains $[g_{\varepsilon}^j]$ was selected that represented $\varepsilon$ newly-formed at that load step (i.e they could not be tracked back to a prior load step).
The detection algorithm, like the duplicate grain detector, utilised checks in both position and orientation space to verify $\gamma \leftrightarrow \varepsilon$ matches.

From grain lists $[g_{\gamma}^i]$ and $[g_{\varepsilon}^j]$, a list of grain pairs $[(g_{\gamma}^i, g_{\varepsilon}^j)]$ was generated.
A separation check for each grain pair was performed, following work by \citet{louca_accurate_2021} -- $g_{\varepsilon}^j$ was considered to be embedded in $g_{\gamma}^i$ if their centre-of-mass separation, $d$, obeyed a function of the austenite grain radius, $d < \chi\left(R_{\gamma}\right)$, where $\chi$ was set to 1.5.
Grain pairs that passed this check were then evaluated to see if they possessed an orientation relationship.

\subsubsection{Orientation relationship detection}
An orientation relationship detection algorithm, Algorithm~\ref{alg:get_matches}, was used to detect orientation relationships between austenite and epsilon grains.
A list of theoretical $\gamma \rightarrow \varepsilon$ rotation matrices was defined from the variants of the Shōji-Nishiyama orientation relationship \citep{shoji_geometrische_1931, nishiyama_x-ray_1934}, as defined below, using the MTEX MATLAB library \citep{bachmann_texture_2010}:
\begin{equation}
    \begin{aligned}
        \hkl{111}_\gamma &\parallel \hkl{0001}_\varepsilon \\
        \hkl<11-2>_\gamma &\parallel \hkl<1-100>_\varepsilon
    \end{aligned}
    \label{eq:SN_OR}
\end{equation}

For each theoretical $\gamma \rightarrow \varepsilon$ rotation matrix from all $\varepsilon$ variants, the misorientation between that and the observed $\gamma \rightarrow \varepsilon$ rotation matrix was determined, considering both cubic and hexagonal symmetries.
The \verb|misorien_from_delta| function and symmetry operators were obtained from the Pymicro Python library \citep{proudhon_pymicro_2021}.
If a misorientation less than \SI{1.5}{\degree} was found, the grain pair $(g_{\gamma}^i, g_{\varepsilon}^j)$ was considered a match.
For each tracked $\gamma$ grain $g_{\gamma}$, the first (i.e earliest load step) appearance of a match to one of its $\gamma$ grain representations $[g_{\gamma}^i]$ was considered to be the point of $\varepsilon$ nucleation within the tracked grain.
The output of Algorithm~\ref{alg:get_matches} was verified against established orientation relationship evaluation routines inside MTEX.

\begin{algorithm}
\caption{A Python function to verify orientation relationships between austenite and epsilon grains.}
\label{alg:get_matches}

\SetKwFunction{CheckForOR}{check\_for\_OR}
\SetStartEndCondition{ }{}{}
\SetKwProg{Fn}{def}{\string:}{}
\SetKw{KwTo}{in}
\SetKwFor{For}{for}{\string:}{}
\SetKwIF{If}{ElseIf}{Else}{if}{:}{elif}{else:}{}

\AlgoDontDisplayBlockMarkers
\SetAlgoNoEnd
\SetAlgoNoLine

\SetKwInOut{Input}{input}
\SetKwInOut{Output}{output}
\SetKwData{gGamma}{aus\_grain}
\SetKwData{gEpsilon}{epsi\_grain}
\SetKwData{variants}{variant\_matrices}
\SetKwData{cubsymms}{cubic\_symmetries}
\SetKwData{hexsymms}{hex\_symmetries}
\SetKwData{desiredmisorien}{desired\_misorien}
\SetKwData{obsmisorien}{observed\_misorien}
\SetKwData{dmisorien}{misorien\_delta}
\SetKwData{misoriendeg}{misorien\_deg}
\SetKwData{misorientol}{misorien\_tol}
\SetKwData{cubsymm}{cubic\_symm\_op}
\SetKwData{hexsymm}{hex\_symm\_op}
\SetKwData{cw}{cw}
\SetKwData{mis}{mis}

\SetKwData{gGammaU}{aus\_grain.U}
\SetKwData{gEpsilonU}{epsi\_grain.U}

\SetKwFunction{misorienfromdelta}{misorien\_from\_delta}
\SetKwFunction{nptrace}{np.trace}
\SetKwFunction{npacos}{np.arccos}
\SetKwFunction{npradtodeg}{np.rad2deg}

\newcommand{\forconde}{\desiredmisorien \KwTo\variants}
\newcommand{\forcondf}{\cubsymm \KwTo\cubsymms}
\newcommand{\forcondg}{\hexsymm \KwTo\hexsymms}

\Fn(){\misorienfromdelta{\dmisorien}}{
    \KwData{\dmisorien: Misorientation matrix}
    \KwResult{\misoriendeg: Angle in degrees}
    \BlankLine

    \cw = (\nptrace{\dmisorien} - 1.0) / 2.0

    \If{\cw $>$ 1}{
        \cw = 1
    }
    \If{\cw $<$ -1}{
        \cw = -1
    }

    \mis = \npacos{\cw}\;
    \misoriendeg = \npradtodeg{\mis}\;

    \Return \misoriendeg
    
}
\BlankLine
\Fn(){\CheckForOR{\gGamma, \gEpsilon, \variants, \cubsymms, \hexsymms, \misorientol}}{
    \KwData{\\\begin{tabular}{rl}
    \gGamma:& $\gamma$ grain $g_{\gamma}^i$\\
    \gEpsilon:& $\varepsilon$ grain $g_{\varepsilon}^j$\\
    \variants:& variant misorientations
            \end{tabular}}
    \KwResult{True/False}
    \BlankLine

    \For{\forconde}{
        \For{\forcondf}{
            \For{\forcondg}{
                \obsmisorien = (\gGammaU $\cdot$ \cubsymm).T $\cdot$ (\gEpsilonU $\cdot$ \hexsymm)\;
                \dmisorien = \desiredmisorien.T $\cdot$ \obsmisorien\;
                \misoriendeg = \misorienfromdelta{\dmisorien}\;
                \If{\misoriendeg $<$ \misorientol}{
                    \Return True
                }
            }
        }
    }
    \Return False
}
\end{algorithm}

\subsubsection{Martensite variant detection}
Martensite variant IDs were determined by iterating over the theoretical $\gamma \rightarrow \varepsilon$ transformation matrices, calculating the predicted $\varepsilon$ orientation from the $\gamma$ orientation, then determining the misorientation angle between theoretical and predicted $\varepsilon$ orientations, taking into consideration the hexagonal symmetry of the $\varepsilon$ grain.
The misorientation angle for each variant was determined, and the final variant ID was taken as the index of the minimum misorientation angle.
Like the orientation relationship detection, the validity of the variant ID calculation routine was verified against established MTEX variant ID routines.

\subsubsection{Minimum strain work prediction}
The prediction of a martensite variant ID using the minimum strain work criterion follows work by \citet{humbert_modelling_2002} and \citep{li_situ_2014}.
A temporary reference frame, $K_t$, was defined using basis vectors given below:
\begin{equation}
\begin{aligned}
\hat{K_t^1} &= \frac{1}{\sqrt{6}}\hkl[11-2]\\
\hat{K_t^2} &= \frac{1}{\sqrt{2}}\hkl[-110]\\
\hat{K_t^3} &= \frac{1}{\sqrt{3}}\hkl[111]\\
\end{aligned}
\end{equation}
In this reference frame $K_t$, the plane strain $\matr{D_{K_t}^{\gamma \rightarrow \varepsilon}}$ that facilitates the shear necessary to transform the FCC $\gamma$ to HCP $\varepsilon$ lattice can be given:
\begin{equation}
\matr{D_{K_t}^{\gamma \rightarrow \varepsilon}} = \begin{bNiceMatrix}[columns-width=auto]
1 & 0 & \frac{\sqrt{2}}{4}\\
0 & 1 & 0\\
0 & 0 & 1\\
\end{bNiceMatrix}
\end{equation}
Rotating this strain matrix to the standard reference frame $K_{\gamma}$ (where the basis vectors $\left(\hat{K_{\gamma}^1}, \hat{K_{\gamma}^2}, \hat{K_{\gamma}^3}\right)$ are the direct lattice vectors) yielded $\matr{D_{K_{\gamma}}^{\gamma \rightarrow \varepsilon}}$:
\begin{equation}
\matr{D_{K_{\gamma}}^{\gamma \rightarrow \varepsilon}} = \frac{1}{12}\begin{bNiceMatrix}[columns-width=auto]
13 & 1 & 1\\
1 & 13 & 1\\
-2 & -2 & 10\\
\end{bNiceMatrix}
\end{equation}
The deformation tensor $\matr{\varepsilon_{K_{\gamma}}^{\gamma \rightarrow \varepsilon}}$ in the standard reference frame could then be defined:
\begin{equation}
\matr{\varepsilon_{K_{\gamma}}^{\gamma \rightarrow \varepsilon}} = \frac{1}{2}\left(\matr{D_{K_{\gamma}}^{\gamma \rightarrow \varepsilon}} + \left[\matr{D_{K_{\gamma}}^{\gamma \rightarrow \varepsilon}}\right]^{\intercal}\right) - \matr{I} = \frac{1}{24}\begin{bNiceMatrix}[columns-width=auto]
2 & 2 & -1\\
2 & 2 & -1\\
-1 & -1 & -4\\
\end{bNiceMatrix}
\end{equation}
However, this deformation matrix only describes a single shear operation leading to a single $\varepsilon$ variant \citep{humbert_modelling_2002}.
\citet{humbert_modelling_2002} specifies that three equivalent deformation matrices exist for each $\varepsilon$ variant, totalling 12 deformation matrices overall.
To generate the 12 deformation tensors $\left.\matr{\varepsilon_{K_{\gamma}}^{\gamma \rightarrow \varepsilon}}\right|_{i}$ given $\matr{\varepsilon_{K_{\gamma}}^{\gamma \rightarrow \varepsilon}}$, both the three-fold rotation symmetry in the $\hkl{111}_{\gamma}$ planes and the four $\gamma \rightarrow \varepsilon$ lattice transformation matrices $\matr{V}^{\gamma \rightarrow \varepsilon}_i$ must be considered.
A \numproduct{3x3x3} hexagonal symmetry operator matrix $\matr{H}$ was defined using the pymicro Python library \citep{proudhon_pymicro_2021}, containing three symmetry operators:
\begin{equation}
    \begin{aligned}
        \matr{H} &= \left[\matr{H_1}, \matr{H_2}, \matr{H_3}\right]\\
        \matr{H_1} &= \begin{bNiceMatrix}[columns-width=auto]
                        1 & 0 & 0\\
                        0 & 1 & 0\\
                        0 & 0 & 1\\
                        \end{bNiceMatrix}\\
        \matr{H_2} &= \begin{bNiceMatrix}[columns-width=auto]
                -\frac{1}{2} & \frac{\sqrt{3}}{2} & 0\\
                -\frac{\sqrt{3}}{2} & -\frac{1}{2} & 0\\
                0 & 0 & 1\\
                \end{bNiceMatrix}\\
        \matr{H_3} &= \begin{bNiceMatrix}[columns-width=auto]
                -\frac{1}{2} & -\frac{\sqrt{3}}{2} & 0\\
                \frac{\sqrt{3}}{2} & -\frac{1}{2} & 0\\
                0 & 0 & 1\\
                \end{bNiceMatrix}
    \end{aligned}
\end{equation}
Using $\matr{H_3}$ and a misorientation matrix for MTEX variant 2, $\matr{V^{\gamma \rightarrow \varepsilon}_2}$, as an example, a rotated deformation tensor $\left.\matr{\varepsilon_{K_{\gamma}}^{\gamma \rightarrow \varepsilon}}\right|_{3,2}$ was calculated:
\begin{equation}
\begin{aligned}
\left.\matr{R}\right|_{3,2} &= \matr{V^{\gamma \rightarrow \varepsilon}_2} \cdot \matr{H^{\intercal}_3} \cdot \matr{V^{\gamma \rightarrow \varepsilon}_1}\\
\left.\matr{\varepsilon_{K_{\gamma}}^{\gamma \rightarrow \varepsilon}}\right|_{3,2} &= \left.\matr{R}\right|_{3,2} \cdot \matr{\varepsilon_{K_{\gamma}}^{\gamma \rightarrow \varepsilon}} \cdot \left.\matr{R^{\intercal}}\right|_{3,2}\\
&= \frac{1}{24}\begin{bNiceMatrix}[columns-width=auto]
2 & 1 & 2\\
1 & -4 & 1\\
2 & 1 & 2\\
\end{bNiceMatrix}
\end{aligned}
\end{equation}
corresponding to variant $V_{\varepsilon4}$ in Table~1 of the work by \citet{humbert_modelling_2002}, which defines variants in a different order to MTEX.
This process was automated in a Python function that calculated the three deformation tensors $\left.\matr{\varepsilon_{K_{\gamma}}^{\gamma \rightarrow \varepsilon}}\right|_{i}$ from $\matr{H}$ given a variant ID and $\matr{\varepsilon_{K_{\gamma}}^{\gamma \rightarrow \varepsilon}}$.
The deformation work $E$ given a deformation tensor $\matr{\varepsilon_{K_{\gamma}}^{\gamma \rightarrow \varepsilon}}$ could then be determined from the matrix double-dot product:
\begin{equation}
E = \frac{1}{2}\matr{\sigma_{K_y}}:\matr{\varepsilon_{K_{\gamma}}^{\gamma \rightarrow \varepsilon}}
\end{equation}
To determine the predicted $\varepsilon$ variant given the orientation of a $\gamma$ grain $\matr{U_\gamma}$ and the stress tensor in the sample reference frame $\matr{\sigma_{K_s}}$, the stress tensor was first rotated into the grain standard reference frame $K_\gamma$, then the deformation work was calculated for each equivalent deformation tensor.
The maximum value across all three deformation tensors was taken for each variant, then the maximum value across all variants was determined, and the corresponding variant ID with the matrix that yielded maximum work was returned.

\subsubsection{Grain neighbourhood identification}
To identify the neighbouring grains ($B_i$) of a given grain $A$, their centre of mass distance, $d$, was evaluated against their radii via the equation below, after \citet{louca_accurate_2021}:
\begin{equation}
d < \phi\left(R_A + R_B\right)
\end{equation}
where $\phi$ is a scaling constant, chosen to be \num{1.5}.

\section{Results}
\subsection{Microstructure}
\begin{figure*}[tp]
    \centering
    \includegraphics[width=\textwidth,center]{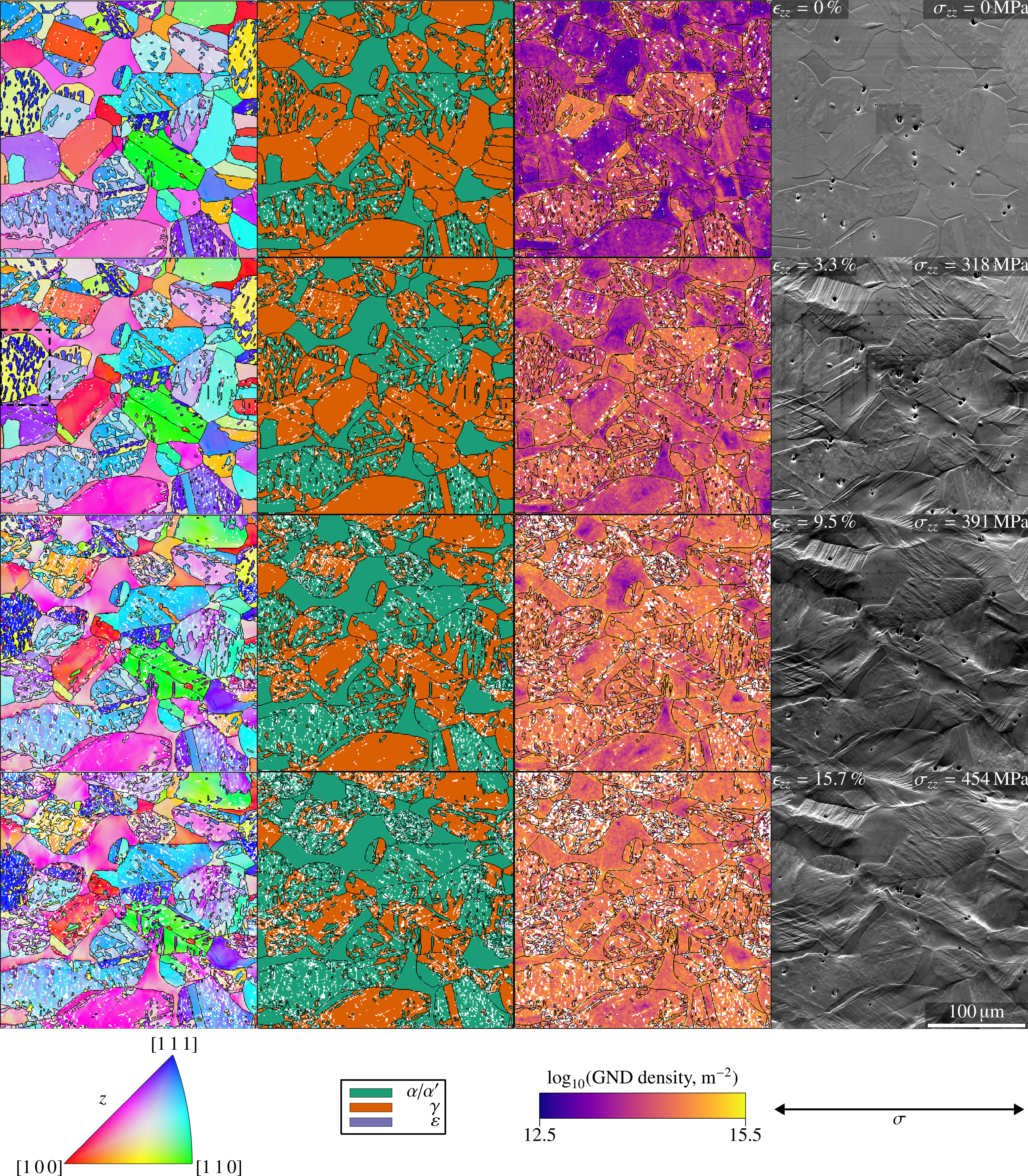}
    \caption{Interrupted-loading EBSD (first three columns) and SE maps (last columns) over increasing applied stress across four load steps (rows). EBSD maps coloured by IPF-$Z$ (first column), phase (second column), logarithmic GND density (third column).}
    \label{fig:ebsd_interrupted_loading}
\end{figure*}

The 304LM steel alloy possesses a complex multi-phase microstructure, as shown in the first row of Figure~\ref{fig:ebsd_interrupted_loading}.

A large fraction of retained austenite ($\gamma$, blue in the phase map) is visible.
Some polygonal ferrite ($\alpha$, red in the phase map) is also present.
The retained austenite grains contain varying amounts of pre-existing martensite ($\alpha'$, red in the phase map).
Although the BCT martensite is indexed as BCC ferrite in the EBSD map, it can be unambiguously distinguished from the polygonal ferrite in the EBSD maps via its increased kernel-averaged misorientation, likely due to the highly twinned or slipped martensitic sub-structure \citep{kelly_1_2012}.
The initial texture of the $\gamma$ and $\alpha$/$\alpha'$ phases from the EBSD map of the first load step is given in Figure~\ref{fig:initial_texture}b.

\subsection{3DXRD}
\subsubsection{Indexing}
In total, \num{72947} grains were indexed over ten load steps.
The breakdown of indexed grain numbers by load step and phase is given in Table~\ref{tab:3dxrd_grains_indexed}.
The error parameters, as determined by the bootstrap process, are given in Table~\ref{tab:error_comparison}, averaged across all recorded raw grains in the experiment.
The tracking data algorithm was able to track a total of \num{10360} grains across more than one load step.

\begin{table}[h]
\centering
\begin{tabular}{cccc}
\hline
\multicolumn{1}{c}{\multirow{2}{*}{\textbf{Applied Load (MPa)}}} & \multicolumn{3}{c}{\textbf{Number of grains indexed}} \\
\multicolumn{1}{c}{} & $\gamma$ & $\alpha$/$\alpha'$ & $\varepsilon$ \\ \hline
0 & 4930 & 3054 & 9 \\
40 & 4888 & 3184 & 15 \\
80 & 4858 & 3242 & 23 \\
120 & 4732 & 3277 & 25 \\
160 & 4607 & 3319 & 59 \\
200 & 4459 & 3323 & 257 \\
240 & 4098 & 3254 & 651 \\
280 & 3068 & 3117 & 910 \\
320 & 1202 & 2688 & 747 \\
36 & 1305 & 2878 & 768 \\ \hline
\end{tabular}
\caption{Number of 3DXRD grains indexed, per phase, at each load step}
\label{tab:3dxrd_grains_indexed}
\end{table}

\begin{table}[h]
\centering
\begin{tabular}{@{}lc@{}}
\toprule
\begin{tabular}[c]{@{}l@{}}Parameter\\ \end{tabular}  & \begin{tabular}[c]{@{}c@{}}Error\\ \end{tabular} \\ \midrule
Orientation (\si{\degree}) & 0.01 \\
Position (\si{\micro\metre}) & 3 \\
$\epsilon$ (diagonal element mean) $\left(\times10^{-3}\right)$ & 0.1\\
$\sigma$ (diagonal element mean, \si{\mega\pascal}) & 32\\ \bottomrule
\end{tabular}
\caption{Mean 3DXRD grain parameter errors in laboratory reference frame.}
\label{tab:error_comparison}
\end{table}

\begin{figure}
    \centering
    \includegraphics[width=\columnwidth]{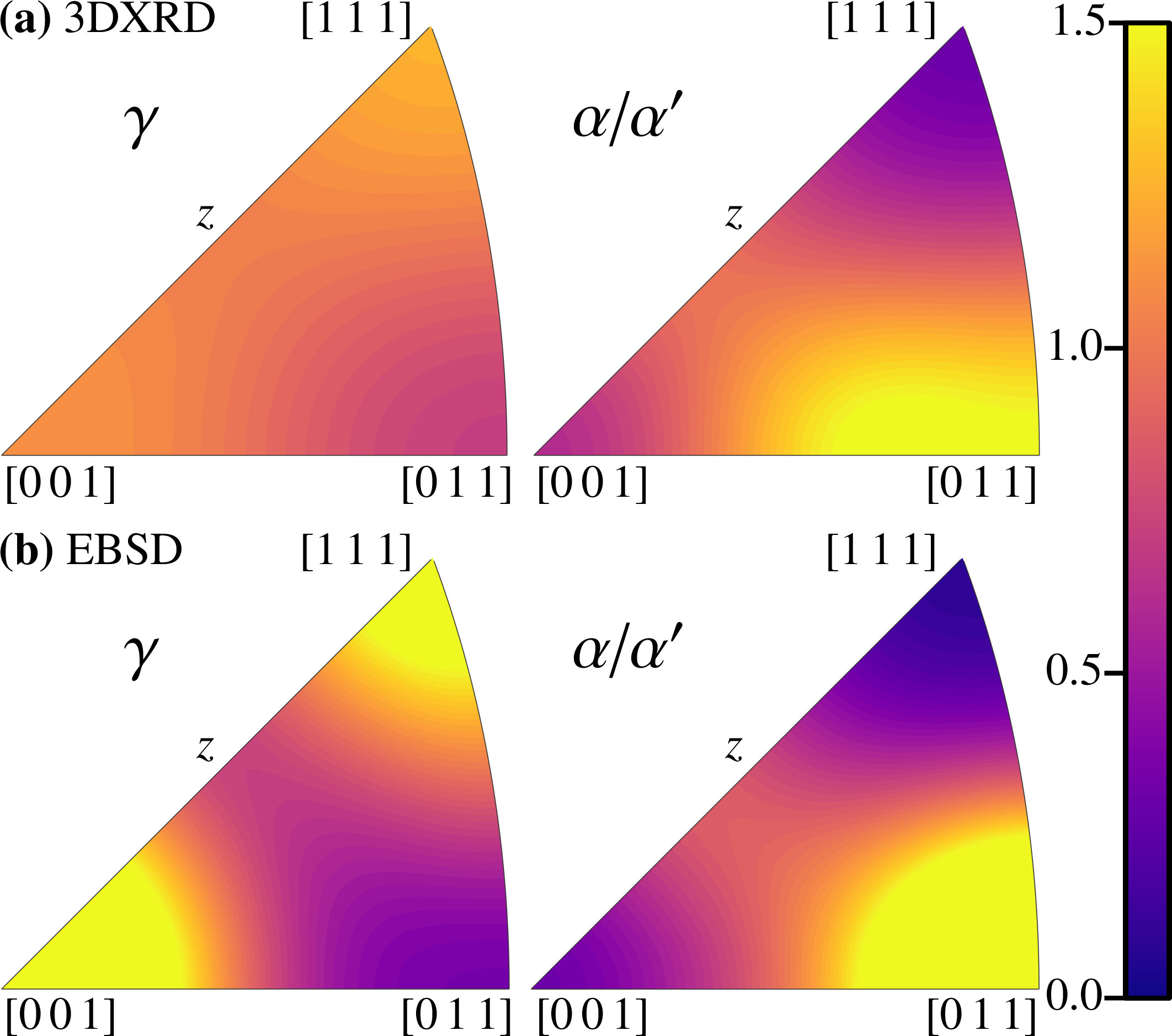}
    \caption{Initial texture of $\gamma$ and $\alpha$/$\alpha'$ from 3DXRD grains (a) and the EBSD map (b) via IPF-$z$ plots.}
    \label{fig:initial_texture}
\end{figure}

\subsubsection{In-situ phase transformation}
\begin{figure}
    \centering
    \includegraphics[width=\columnwidth]{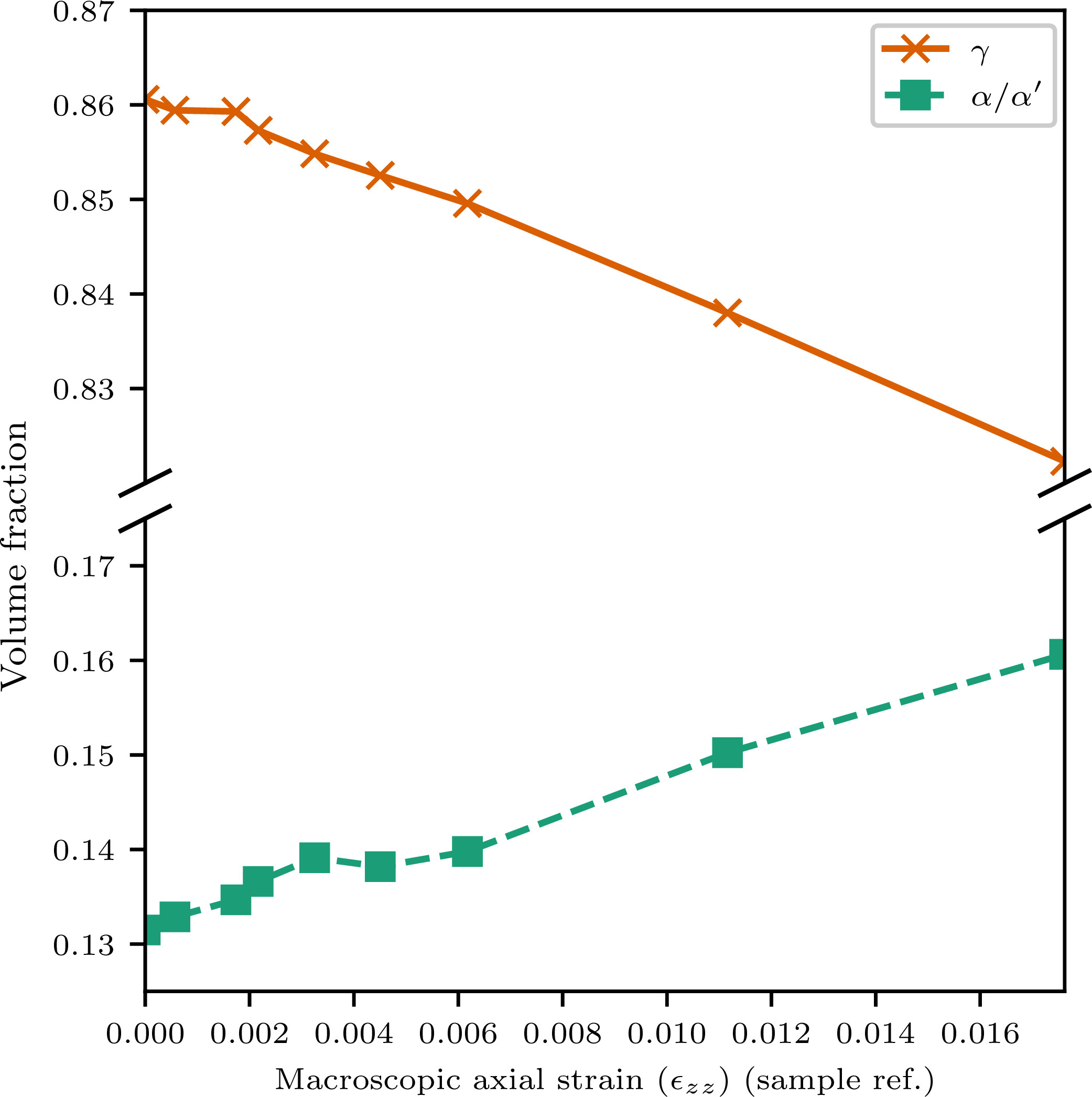}
    \caption{Change in Rietveld volume fractions of austenite ($\gamma$) and ferrite/martensite ($\alpha$/$\alpha'$) against applied stress.}
    \label{fig:3dxrd_phase_fractions}
\end{figure}

The phase evolution as a function of applied stress, as determined by the Rietveld refinement, are shown in Figure~\ref{fig:3dxrd_phase_fractions}.

\begin{figure*}[tp]
    \centering
    \includegraphics[width=140mm,center]{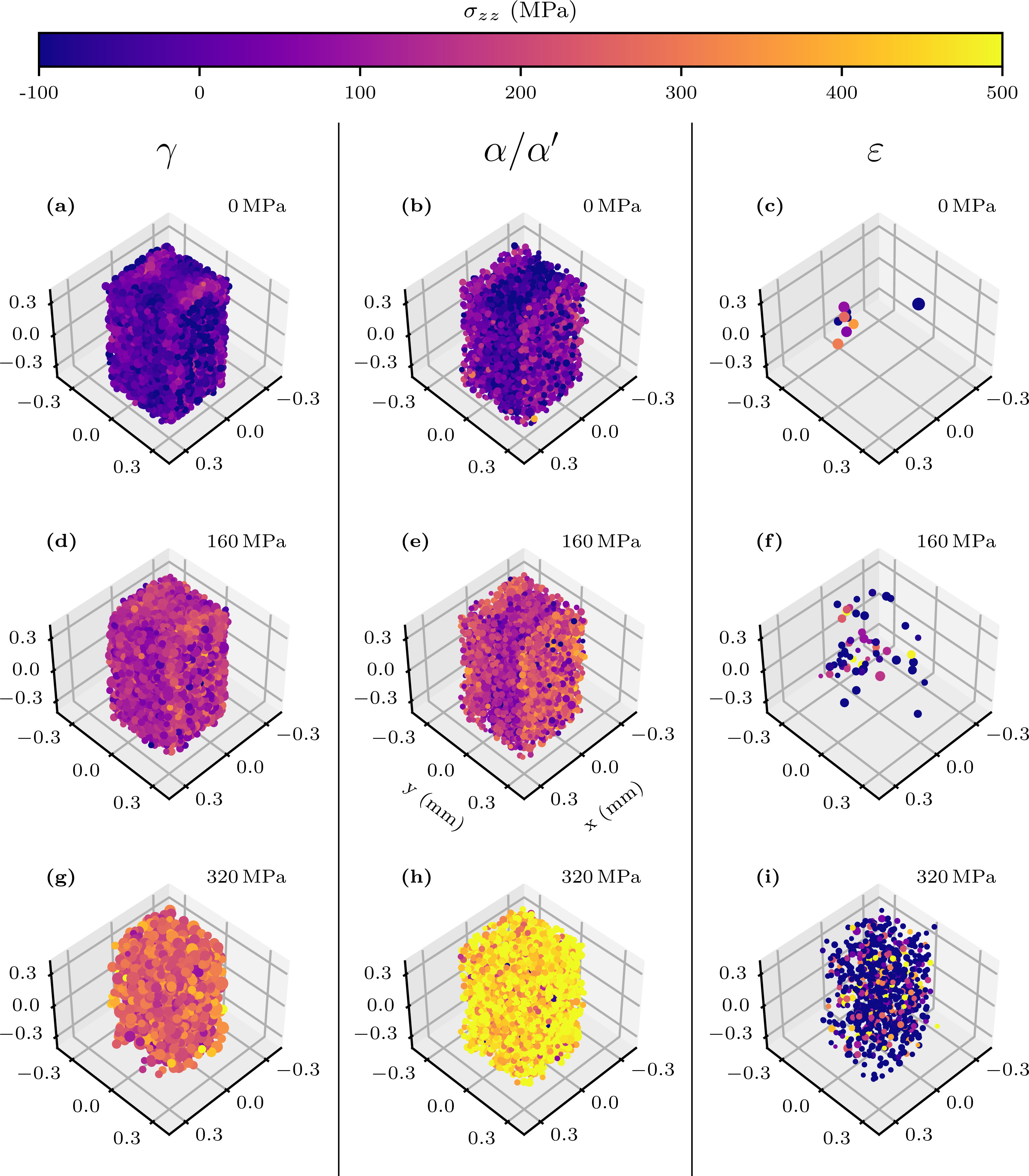}
    \caption{Grain maps of austenite, $\gamma$,  (a, d, g), ferrite/martensite, $\alpha$/$\alpha'$ (b, e, h) and epsilon, $\varepsilon$ (c, f, i) phases at three different applied stress levels, coloured by grain-level axial stress, $\sigma_{zz}$.}
    \label{fig:grain_maps}
\end{figure*}

The initial balance of phases comprises predominantly austenite, with $\alpha$/$\alpha'$ making up \qty{\sim 13}{\percent} of the volume.
A very small amount (\qty{< 5}{\percent}) of $\varepsilon$ is present in the as-received condition -- an exact volume fraction cannot be given here due to the errors associated with the Rietveld quantitative phase assessment.
These starting proportions remain approximately constant until around \num{0.005} macroscopic strain, where the volume fraction of $\gamma$ starts to decrease, coupled with an increase in $\alpha$/$\alpha'$.
Maps of grain positions across a selection of load steps, coloured by grain axial ($zz$) stress in the sample reference frame, for each phase, are provided in Figure~\ref{fig:grain_maps}.

\begin{figure}[ht!]
    \centering
    \includegraphics[width=\columnwidth]{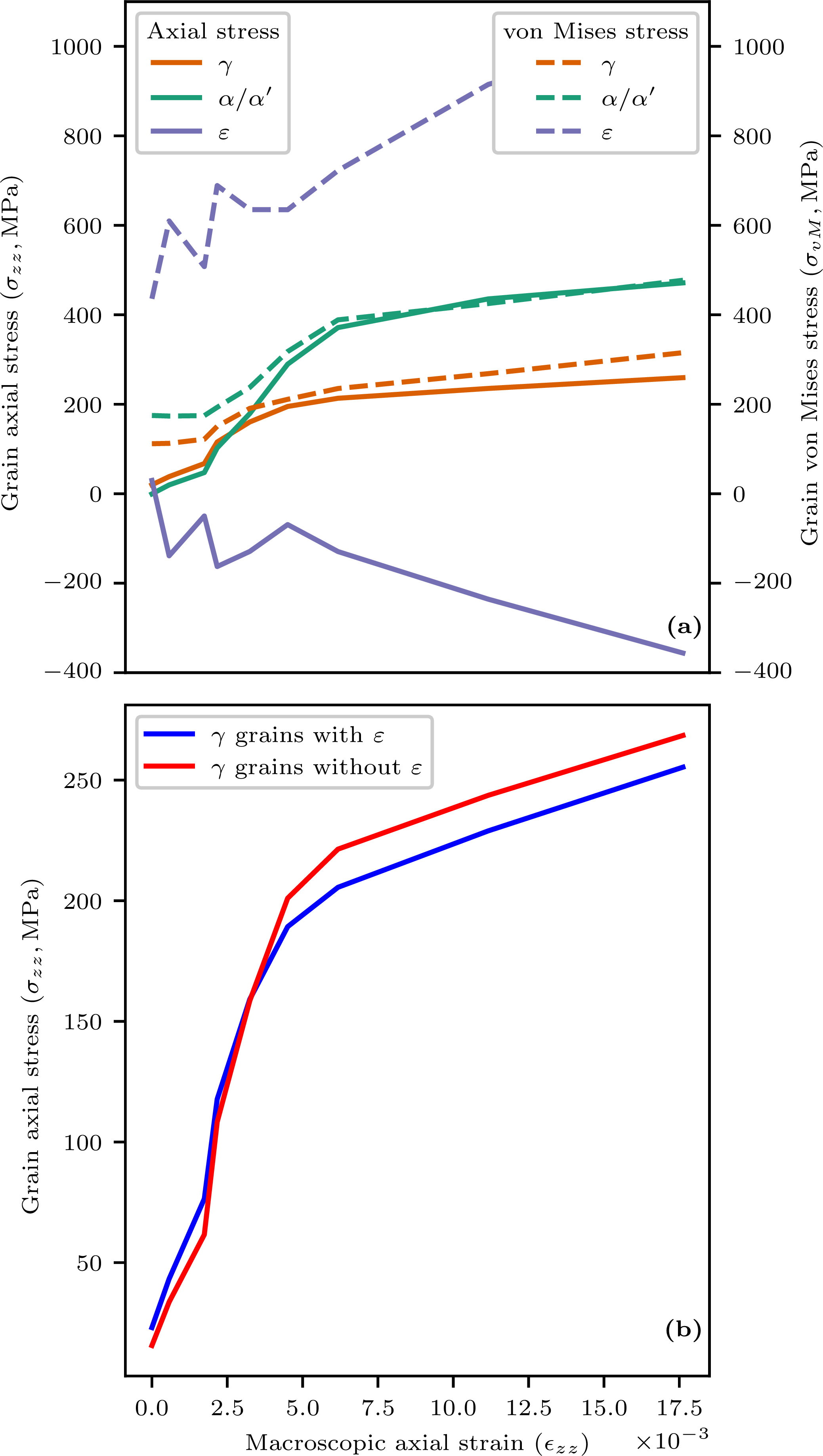}
    \caption{Development of grain $zz$ stress (\textbf{(a)} and \textbf{(b)}, left axis) and von~Mises stress (\textbf{(a)}, right axis) vs applied strain.}
    \label{fig:grain_stress_development}
\end{figure}

A small amount of residual tensile stress can be seen, particularly on the $xz$ surfaces of the $\alpha$/$\alpha'$ map ((b) and (e)).
These residual stresses are not observed to affect the nucleation of $\varepsilon$, as no significant bias is observed in the positions of $\varepsilon$ grains.
The grain axial stresses in the $\gamma$ and $\alpha$/$\alpha'$ phases are observed to increase as the applied load increases.
The stress development of the $\varepsilon$ phase is more complex.
To investigate this further, the volume-weighted mean grain axial stresses as a function of applied strain for each phase are plotted in Figure~\ref{fig:grain_stress_development}a.
The mean stress in each phase was close to zero with no load applied, indicating a good choice of reference lattice parameters for each phase.
The stress in $\alpha$/$\alpha'$ was observed to increase significantly beyond than the stress in the $\gamma$ phase, indicating substantial stress partitioning to the BCC phase, which has been observed in other XRD studies of metastable steels \citep{hidalgo_interplay_2019}.
The $\varepsilon$ phase, in this axial direction, is subjected to substantial compressive stresses.
Given its volume fraction is low, a strong stress partitioning effect is not expected in the parent $\gamma$ phase.
By also plotting the von~Mises stress state of each grain, the negative nature of the $\varepsilon$ axial stress can be explored.
The stress state of the $\gamma$ and $\alpha$/$\alpha'$ phases is largely dominated by the applied stress, especially beyond the approximate global yield point.
The stress state of the $\varepsilon$ phase, however, is highly variable and must therefore be strongly influenced by local crystallographic effects associated with its creation and interface with the surrounding $\gamma$.

\begin{figure}
    \centering
    \includegraphics[width=\columnwidth]{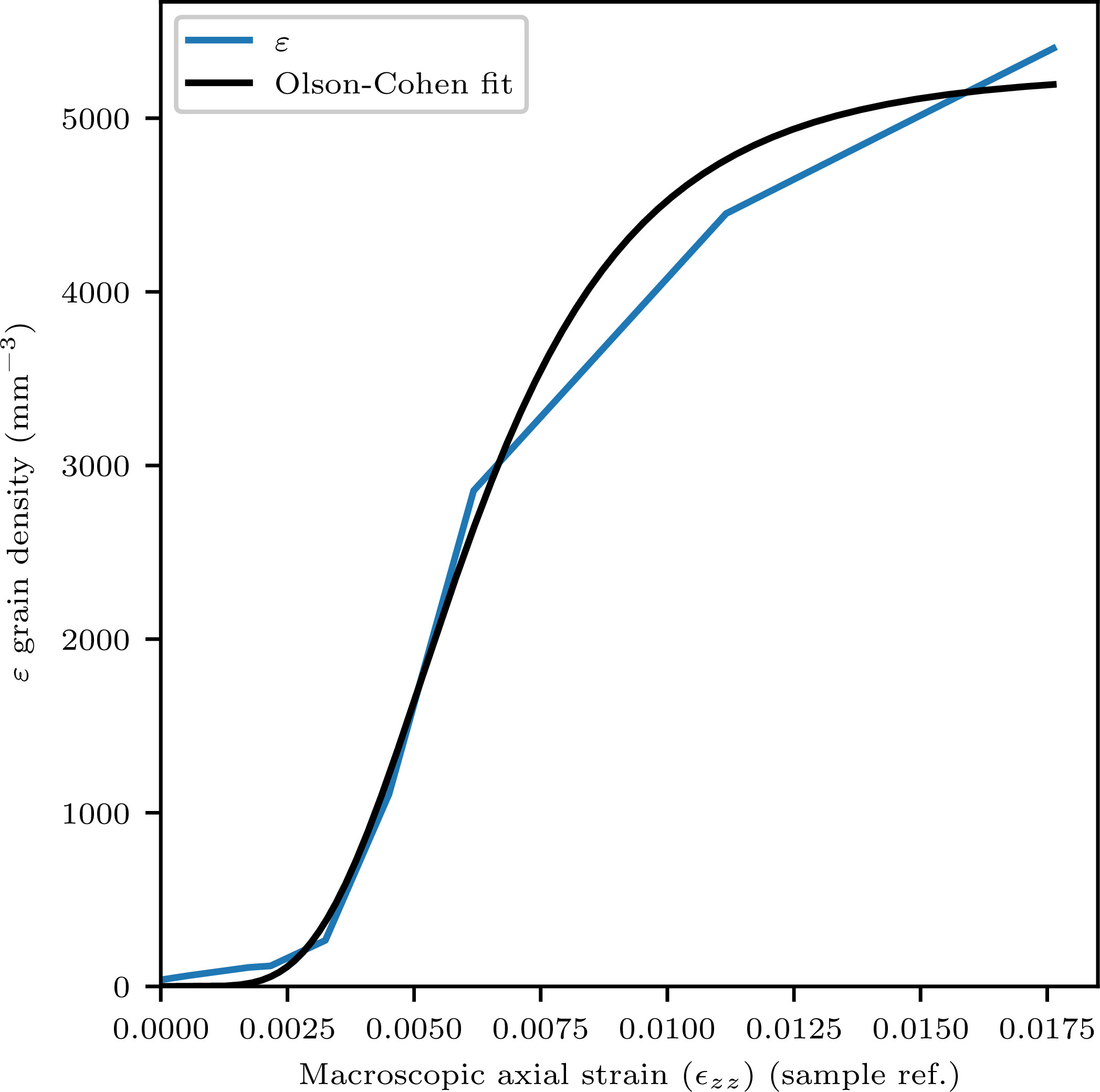}
    \caption{Epsilon grain nucleation with increasing applied strain, and Olson-Cohen kinetics equation fit.}
    \label{fig:n_epsilon_increase}
\end{figure}

To evaluate the $\varepsilon$ nucleation kinetics, the number of new $\varepsilon$ grains appearing at each load step (as determined by the tracking algorithm) was plotted against the macroscopic axial strain in Figure~\ref{fig:n_epsilon_increase}.
The scipy Python library \citep{virtanen_scipy_2020} was used to fit a modified form of the Olson-Cohen equation (given below) \citep{olson_kinetics_1975} connecting the number of $\varepsilon$ martensite grains $n_{\varepsilon}$ to the degree of plastic strain applied:
\begin{equation}
    n_{\varepsilon} = s\left(1-\exp{\left\{-\beta\left[1-\exp{\left(-\alpha\epsilon\right)}\right]^{n}\right\}}\right)
\end{equation}
where $s$ is a scale factor, and all other terms have their original Olson-Cohen meanings \citep{olson_kinetics_1975}.
The scipy-optimized values for $s$, $\alpha$, $\beta$, $n$ were \num{1500}, \num{309}, \num{1.76}, and \num{7.38} respectively.
The result of this fit is also plotted in Figure~\ref{fig:n_epsilon_increase}.
In total, \num{935} unique $\varepsilon$ grains could be matched to the tracked $\gamma$ grains using their positions and orientations.

\subsubsection{Correlation with grain diameter}
\begin{figure}
    \centering
    \includegraphics[width=\columnwidth]{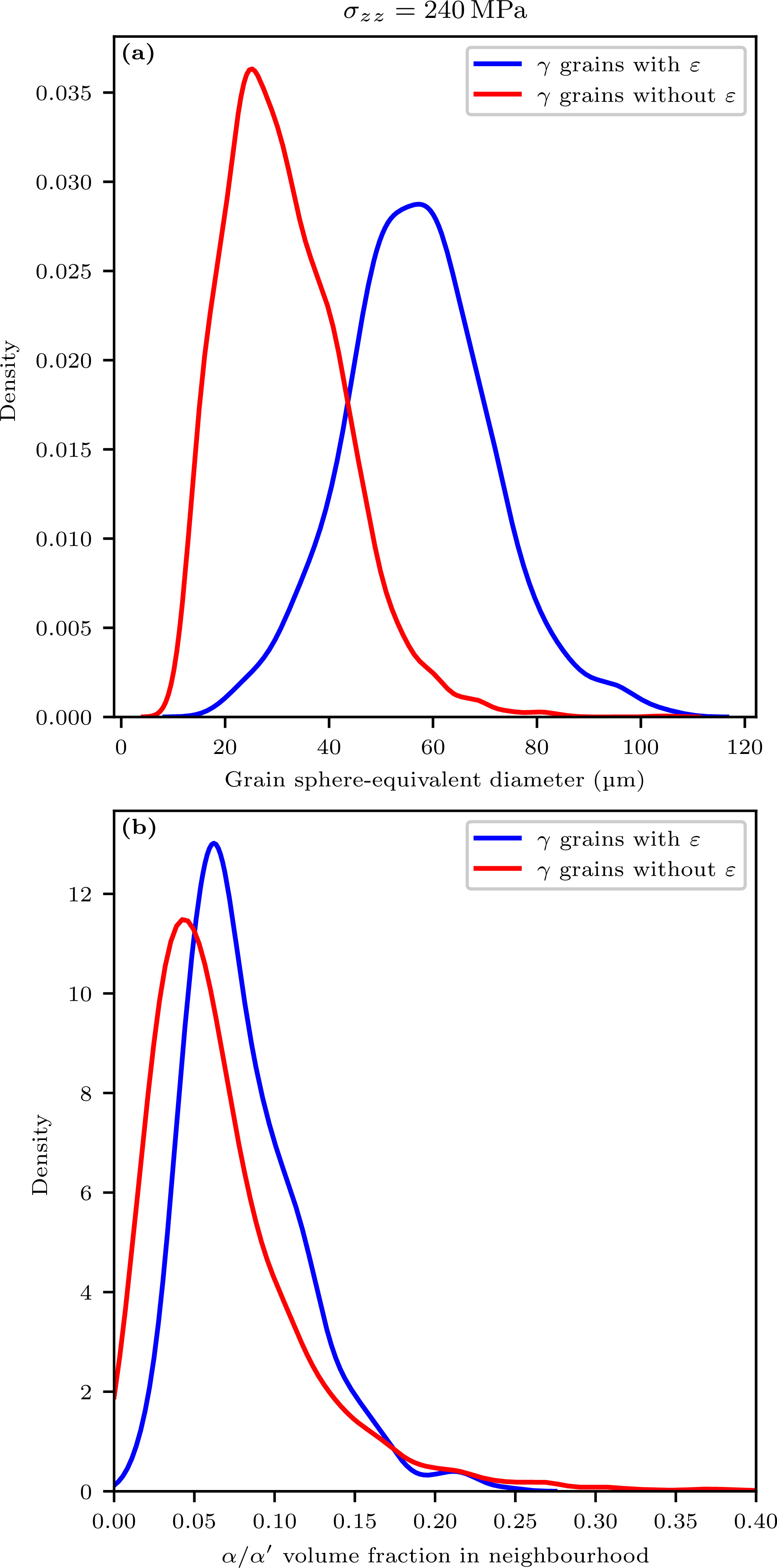}
    \caption{Comparison of distributions of austenite grain diameter (a) and ferrite/martensite density of neighbourhood (b) with and without a detected $\varepsilon$ transformation at an applied stress of \qty{240}{\mega\pascal}.}
    \label{fig:with_without_dists}
\end{figure}

With the $\gamma$ and $\varepsilon$ grains associated, the stability of the $\gamma$ grains could then be probed, using the formation of $\varepsilon$ within them as a measure of the $\gamma$ resistance to transformation.
All tracked $\gamma$ grains that appeared at an applied stress of \qty{240}{\mega\pascal} were extracted, and divided into two groups depending on whether or not $\varepsilon$ had been detected by that point within said grains.
Histograms of the diameters of these two groups, plotted in Figure~\ref{fig:with_without_dists}a, show a clear difference in distribution, where grains containing $\varepsilon$ were much larger on average than grains without.
The statistical significance of this was confirmed with a Student's t-test utilising the scipy Python library \verb|stats.ttest_ind| function, yielding $p < 0.001$.
This suggests that larger $\gamma$ grains are less intrinsically stable against transformation than smaller $\gamma$ grains.

\subsubsection{Correlation with grain neighbourhood}
To investigate whether austenite grain stability was connected to the configuration of its immediate neighbourhood, the directly adjacent grains were identified for every tracked $\gamma$ grain, with or without $\varepsilon$, as appeared at \qty{240}{\mega\pascal}.
The neighbours of each tracked $\gamma$ grain were split into austenitic and\\ ferritic/martensitic neighbours, and their volumes were used to compute a $\alpha$/$\alpha'$ neighbourhood volume fraction value.
The histograms of these values for both $\varepsilon$-containing and $\varepsilon$-lacking $\gamma$ grains are plotted in Figure~\ref{fig:with_without_dists}b.
From the distributions plotted, it is evident that $\gamma$ grains with a lower stability tended to have a greater ferritic/martensitic volume fraction in the local grain neighbourhood.
A t-test was again employed to determine the statistical significance, yielding $p < 0.001$.
No statistically significant correlation between neighbourhood grain orientation and austenite grain stability was observed.

\subsection{Interrupted-loading EBSD}
\begin{figure}
    \centering
    \includegraphics{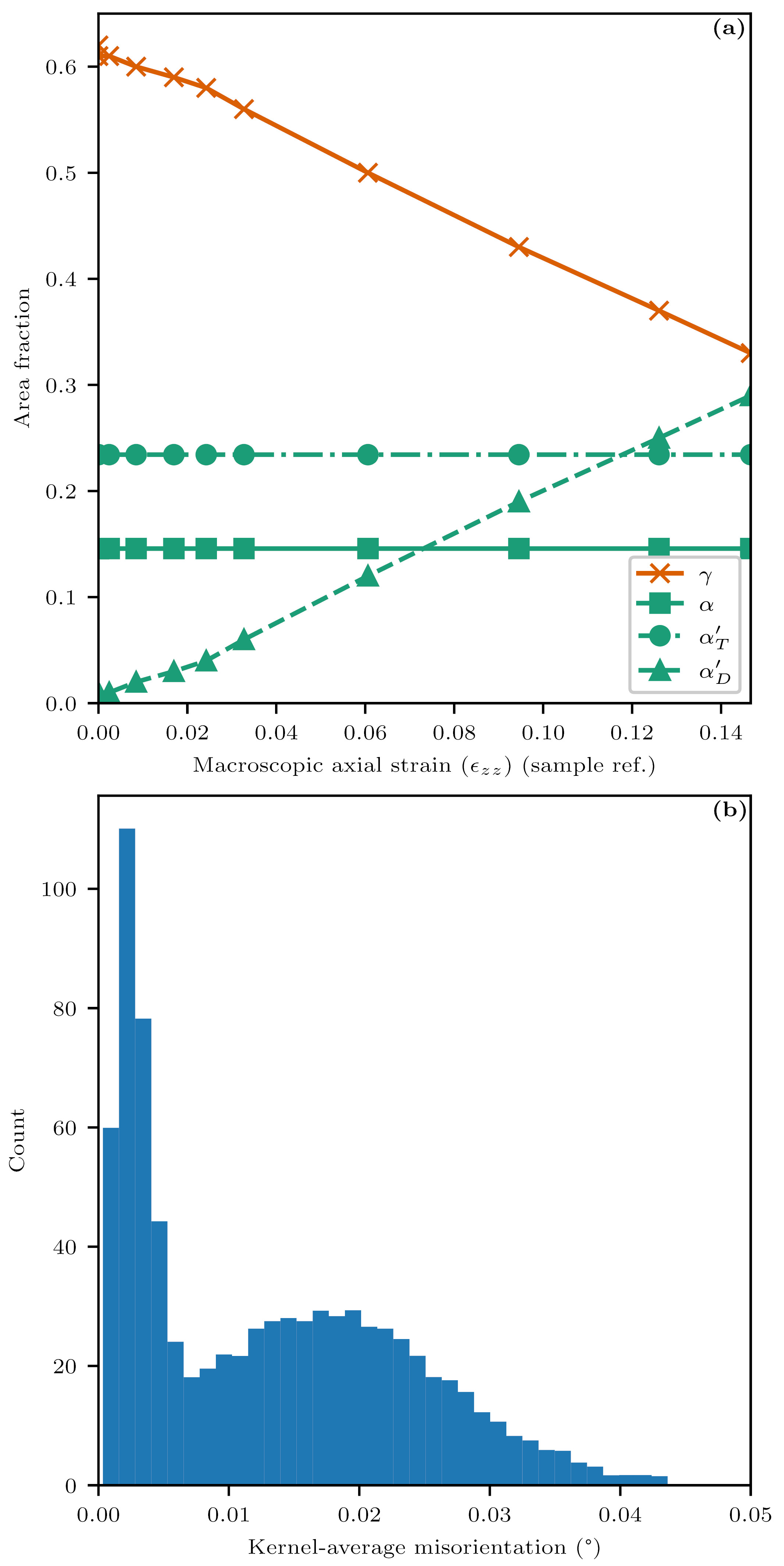}
    \caption{EBSD phase fraction development across increasing deformation (a) and kernel-average misorientation distributions for EBSD BCC phase (b).}
    \label{fig:ebsd_phase_fractions_and_kam}
\end{figure}

A total of 11 load steps were studied; for each, an interrupted-loading EBSD measurement was obtained.
A selection of the EBSD maps collected at four load steps is shown in Figure~\ref{fig:ebsd_interrupted_loading}.
At each load step, the orientation map (by IPF-$z$), phase map, GND density, and secondary electron map is provided.
The DIMT is clearly visible, with a significant portion of most $\gamma$ grains transforming to $\alpha'$, as shown in Figure~\ref{fig:ebsd_phase_fractions_and_kam}a.
From the GND maps, a clear difference is seen between the GND values in the polygonal ferrite and athermal martensite.
This is more clearly apparent in the kernel-averaged misorientation (KAM) data -- taking a histogram of these values across all BCC-indexed pixels of the as-received EBSD map yielded Figure~\ref{fig:ebsd_phase_fractions_and_kam}b.
The distinction in KAM values of polygonal ferrite (the sharp low-angle peak) and athermal martensite (the broad higher-angle peak) is clearly visible.
This allowed the area fraction of polygonal ferrite, $\alpha$, and athermally-produced martensite, $\alpha'_{T}$, to be calculated, which were assumed constant under deformation, thereby allowing the changing fraction of deformation-induced ferrite, $\alpha'_{D}$, to be determined.
Deformation-induced martensite is observed to grow both via the growth of pre-existing "blocky" $\alpha'$ (e.g the yellow grain on the left-hand edge of the IPF-$z$ map), and by the formation of thin laths from shear bands (the grain just below it), which is associated with substantial surface relief, as is evident in the secondary electron images.
The formation of these surface bands can also be seen on the GND maps, particularly in the large pink grain along the bottom edge of the IPF-$z$ map.

\section{Discussion}
\subsection{As-received alloy condition}
\subsubsection{Microstructure}
It is clear from the EBSD maps present in Figure~\ref{fig:ebsd_interrupted_loading} that the as-received microstructure is complicated and requires careful description.
Large grains of polygonal ferrite are present, indicating that the ferrite "nose" was hit during cooling, either due to inaccuracies in composition, simulation of the cooling curves with CALHPAD software, or an insufficient cooling rate during the air quench.
Due to the high solution annealing temperature and time used (\qty{1240}{\celsius} for \qty{12}{\hour}), it is unlikely that the sample was not fully austenitized during this process.
The sample also contained \qty{\sim 20}{\percent} martensite at room temperature -- it was possible to separate polygonal ferrite from this martensite (see Figure~\ref{fig:ebsd_phase_fractions_and_kam}a) using the kernel-averaged misorientation, due to the highly disordered internal structure of the martensite.
The "blocky" morphology of the martensite suggests that it likely formed athermally as a result of the air quench -- prior observations by \citet{naraghi_spontaneous_2011, tian_comparing_2018} support this claim and suggest grain boundaries as the primary nucleation site of this "autocatalyic" martensite; this is evident in this alloy.
The additional presence of polygonal ferrite also suggests that the discrepancy between expected (fully austenitic) and observed microstructures is likely due to compositional variability during the alloy production.
This martensite is unlikely to be induced by mechanical preparation of the sample due to the final electrolytic polishing step which removed the deformed layer induced by the mechanical polishing alone \citep{takeda_mechanism_2021}.
A search of the literature for alloys with similar morphologies yields a study by \citet{naraghi_spontaneous_2011} who found similar large "blocky" athermal martensite in a 301 stainless steel (17Cr, 7Ni) similar to the steel composition used in this study (19Cr, 7Ni) that had been quenched to LN2 temperatures.
These observations suggest an extremely unstable retained austenite phase, and although some of the less stable austenite had already transformed into athermal martensite, it is clear from the phase fraction development in Figure~\ref{fig:ebsd_phase_fractions_and_kam}a that substantial further transformations did occur.
A large difference in phase balance is observed between 3DXRD (Figure~\ref{fig:3dxrd_phase_fractions}) and EBSD (Figure~\ref{fig:ebsd_phase_fractions_and_kam}a) datasets, which is attributed to variability between samples - this discrepancy was also reported in prior measurements of this alloy with diffraction-contrast tomography (DCT) \citep{ball_registration_2023}.

A lack of $\varepsilon$ presence in the initial 3DXRD phase balance suggests no substantial deformation-induced transformation in the initial condition for the 3DXRD sample.
The initial grain-averaged axial stresses being approximately zero for each phase (as per Figure~\ref{fig:grain_stress_development}a) suggests good choices of reference lattice parameter for all phases.
The challenges faced with separating athermal martensite from polygonal ferrite also apply to the 3DXRD data.
Like in the EBSD maps, the BCT martensite and the BCC polygonal ferrite will index the same way in the 3DXRD indexing pipeline.
Unlike the EBSD maps, the differences between the BCT and BCC phases in grain-averaged far-field 3DXRD data are extremely difficult to discern.
Like martensite, polygonal ferrite is expected to exhibit the same orientation relationships as martensite to the parent austenite phase \citep{sharma_preferential_2016}, ruling out this property as a means to distinguish between the two phases.
These difficulties, coupled with the lack of substantial $\alpha'$ volume fraction growth in the \qty{2}{\percent} strain region measured with the 3DXRD technique, explains why deformation-induced $\alpha'$ was not explored in the 3DXRD data as part of this study.

\subsubsection{Texture}
Figure~\ref{fig:initial_texture} shows the initial texture for the $\gamma$ and $\alpha$/$\alpha'$ phases using both EBSD and 3DXRD techniques.
A mild texture in the $\gamma$ phase is seen in the 3DXRD data, although this is much stronger in the EBSD data, likely due to the small number of grains measured with EBSD.
The $\alpha$/$\alpha'$ phases have a strong \hkl[110]||TD rolled texture in both 3DXRD and EBSD plots, supporting prior measurements of this alloy \citep{ball_registration_2023}, indicating no significant orientation bias in the 3DXRD dataset.

\subsection{3DXRD indexing}
The initial 3DXRD results presented in Table~\ref{tab:3dxrd_grains_indexed} demonstrate a very large total number of grains indexed (\num{72947} across all load steps).
This is a significant undertaking for a phase-transforming alloy -- very few prior studies of the DIMT in steels using 3DXRD have been found in the literature, with most recent work being performed by \citeauthor{hedstrom_deformation_2005} and \citeauthor{neding_stacking_2021} \citep{hedstrom_deformation_2005, hedstrom_stepwise_2007, hedstrom_elastic_2008, neding_formation_2021, neding_stacking_2021}, usually on a very small number of grains.
The grain tracking procedure reduced this number of grains to \num{10360}, demonstrating an accurate sample reference frame alignment and grain tracking methodology.

The maps of grain centre-of-mass position, as shown in Figure~\ref{fig:grain_maps}, show a high degree of convergence in the grain parameter refinement stage of the indexing pipeline, reflecting a good choice of indexing parameters for all phases indexed.
The macroscopic cross-sectional shape of the sample is clearly visible, and the density appears uniform.
A substantial decrease in the number of $\gamma$ grains between the first and last load step is evident -- this is likely due to the early stages of $\alpha'$ nucleation, which will introduce significant strain fields into the parent austenite phase (as can be seen from the GND maps in Figure~\ref{fig:ebsd_interrupted_loading}), making 3DXRD indexing more difficult.
The 3DXRD errors presented in Table~\ref{tab:error_comparison} are more than suitable for the analysis performed in this paper, and are very similar in magnitude to prior far-field 3DXRD studies performed at the ID11 beamline which characterise their errors using different software approaches \citep{oddershede_determining_2010}.

\subsection{Deformation-induced transformation}
It is clear from the increase in the number of unique $\varepsilon$ grains with increasing deformation (Figure~\ref{fig:n_epsilon_increase}) that a deformation-induced phase transformation was induced within the strain range limits imposed by the 3DXRD technique.
This is a major success for this study and indicates a very low degree of austenite stability, as theorised from the as-received microstructural analysis.
Prior studies that have directly analysed $\varepsilon$ formed under in-situ deformation with XRD are rare \citep{hedstrom_elastic_2008, ullrich_competition_2021, li_situ_2014, gauss_situ_2016, tian_micromechanics_2018, barriobero-vila_deformation_2021}, and 3DXRD studies of $\varepsilon$ are rarer still \citep{hedstrom_elastic_2008} and consider only a few grains, compared to the \num{935} unique $\varepsilon$ grains measured with this study.
Nevertheless, the evaluation of the Olson-Cohen model in Figure~\ref{fig:n_epsilon_increase} employed $\varepsilon$ grain number as a proxy for volume fraction -- the relatively low volume fraction of $\varepsilon$ seen in the Rietveld refinement results in Figure~\ref{fig:3dxrd_phase_fractions} makes accurate quantitative phase assessment of this phase difficult, a problem well-known to the Rietveld refinement community \citep{zhao_error_2018}. 

\begin{figure*}[t]
    \centering
    \includegraphics[width=\textwidth,center]{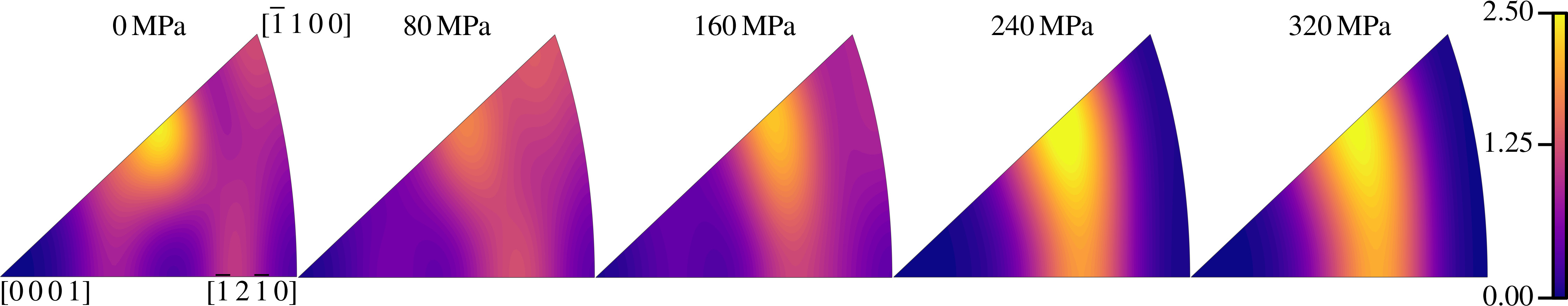}
    \caption{Evolution of $\varepsilon$ texture over increasing applied load via IPF-$z$ plots.}
    \label{fig:eps_texture_evolution}
\end{figure*}

\begin{figure}
    \centering
    \includegraphics[width=\columnwidth]{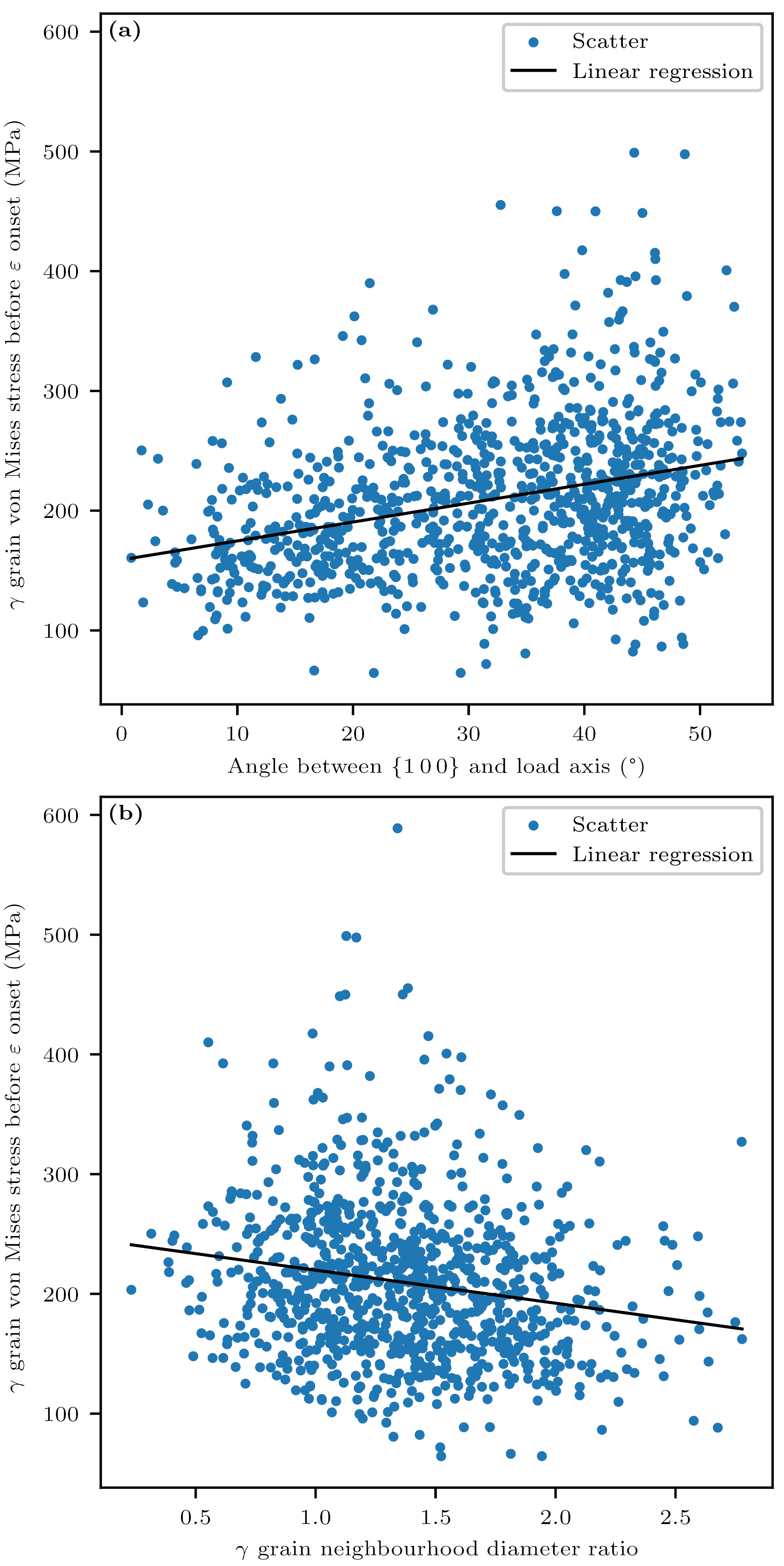}
    \caption{von~Mises stress before transformation vs Austenite grain minimum angle between \hkl{100} and loading axis (a) and ratio of austenite grain diameter to mean neighbourhood grain diameter (b)}
    \label{fig:100_angle_vs_vm}
\end{figure}

The development of the $\varepsilon$ texture under applied load was determined and is shown in Figure~\ref{fig:eps_texture_evolution}.
A clear texture effect is observed, with \hkl{10-13} forming parallel to the loading axis, which, to the author's knowledge, has only been observed once before using synchrotron X-rays \citep{ullrich_competition_2021}, although it has also been observed with EBSD \citep{ueji_crystallographic_2013}.
\citet{ullrich_competition_2021} attributes this texture to the preferential transformation of $\gamma$ grains with \hkl{100} oriented parallel to the loading axis - this mirrors in-situ observations of the DIMT by \citet{neding_stacking_2021, blonde_high-energy_2012} who found decreased stability for austenite grains oriented in this way.
To investigate this further, the angle between the \hkl{100} axis and the loading axis can be plotted against the grain-level von~Mises stress just before transformation -- grains less stable against transformation should require a lower overall applied stress to transform.
Figure~\ref{fig:100_angle_vs_vm}a shows the result of this -- a clear statistically significant ($p < 0.001$) correlation is observed, with grains oriented close to $\hkl{100} \parallel z$ tending to have a lower von~Mises stress just before $\varepsilon$ formation, demonstrating a clear effect of austenite orientation on stability against transformation.

From the EBSD data, $\alpha'$ growth is observed to occur from existing $\alpha'$ colonies and grain boundaries.
From the SEM images, two different morphological forms of transformation-induced $\alpha'$ can be seen -- some $\gamma$ grains, such as the grain in the middle of the left-hand edge (yellow with blue martensite in the IPF-$z$ plot of the first load step), are almost entirely transformed to $\alpha'$ over the course of the loading cycle.
However, there is little change to the surface profile of this grain, as can be seen in SEM images (see Figure \ref{fig:ebsd_interrupted_loading}).
In contrast, other grains, such as the deep orange grain above the grain previously mentioned, form very thin martensite laths, which cannot be indexed by the EBSD process during the early stages of deformation, but are clearly visible via the surface relief in the SEM images.
These different modes are attributed to the range of initial morphologies present in the as-received condition of the alloy.
Pre-existing "blocky" $\alpha'$ in these highly unstable $\gamma$ grains appears to promote the further growth of $\alpha'$ in its surrounding $\gamma$ grain, while suppressing substantial lath $\alpha'$ that would cause a clear surface profile effect, which is supported by prior literature \citep{naraghi_spontaneous_2011, tian_comparing_2018}.

The polygonal ferrite grains, clearly distinguished by their lack of transformation, are significantly affected by martensite transformation in the surrounding grains. This can be most clearly seen in the higher-resolution post-mortem EBSD maps in Figures~\ref{fig:load_12_ebsd_ipf}~and~\ref{fig:load_12_ebsd_kam} of the Supplementary Materials.
Substantial orientation gradients are observed, indicating increased accumulated plastic deformation, particularly in regions of the polygonal ferrite grains close to $\gamma$ grains with substantial martensite transformations.
Prior literature supports these findings, with work by \citet{kang_digital_2007, marteau_investigation_2013, kang_situ_2023, tasan_strain_2014} indicating strain localisation in ferrite grains that neighbour transformed martensite grains.
This is clear evidence of a neighbourhood effect caused by the martensitic transformation itself.
The inverse effect (of grain neighbourhoods on the transformation) is also explored in Section~\ref{sec:dis:neighbourhood}.

\subsubsection{Effect of DIMT on grain stresses}
By connecting $\varepsilon$ grains to their $\gamma$ parents, the effect of the $\varepsilon$ transformation on the individual stress states of parent grains can be determined.
As shown in Figure~\ref{fig:grain_stress_development}b, $\gamma$ grains with $\varepsilon$ had lower axial stresses on average than grains without $\varepsilon$, an effect that only manifested after the approximate macroscopic yield strain of \num{0.005} as seen in Figure~\ref{fig:3dxrd_geom_and_stess_strain}b.
Although this is a small effect, roughly on the order of the stress error, it persists beyond the macroscopic yield point until the highest applied load.
This is explained by the significantly-harder $\varepsilon$ phase, accumulating high levels of elastic stress compared to its parent $\gamma$ grain.
This phenomena is supported by prior literature \citep{lai_deformation-induced_2020, liang_microstructural_2009}.
Using $\varepsilon$ formation as a direct measurement of $\gamma$ grain stability has enabled the effect of a number of micromechanical phenomena on austenite stability to be explored, which will hereby be discussed.

\subsection{Influences on austenite grain stability}
The influence of austenite grain orientation on stability has been clearly demonstrated in Figure~\ref{fig:100_angle_vs_vm}a.
The influence of other microstructural properties such as grain size and neighbourhood configuration will now be explored.

\subsubsection{Grain size}
As shown in Figure~\ref{fig:with_without_dists}a, there is a clear difference in grain size distribution between grains with and without $\varepsilon$, where grains with $\varepsilon$ appear significantly larger.
The relationship between $\gamma$ grain size and stability has been a point of significant contention in the literature.
Many modelling and experimental studies demonstrate that larger austenite grains are less stable, and that refining the grain size increases stability against transformation \citep{turteltaub_grain_2006, haidemenopoulos_kinetics_2014, jung_effect_2011, jimenez-melero_martensitic_2007} for a range of grain sizes from \qtyrange[range-phrase=--,range-units=single]{0.1}{20}{\micro\metre}.
However, select studies have shown the opposite effect, for both very small grains \qty{> 1}{\micro\metre} in ultra-fine-grained austenitic stainless steels \citep{somani_enhanced_2009, marechal_linkage_2011, kisko_influence_2013}, grains between \SI{2.5}{\micro\metre} in size \citep{toda_multimodal_2022}, and very large grains \qtyrange[range-phrase=--,range-units=single]{52}{284}{\micro\metre} \citep{shrinivas_deformation-induced_1995}.
Our data, in contrast, exist somewhere in-between these ranges, but shows a clear statistically-significant difference in size distributions between grains with and without a detected $\varepsilon$ formation, supporting the idea that larger grains are less stable than smaller grains, for the grain size range presented in this study.
This conclusion must be taken with some scrutiny - larger $\gamma$ grains can form larger $\varepsilon$ grains within them, as the length of the $\varepsilon$ is restricted by the grain diameter.
Therefore, the detection limits must be considered, to determine the critical $\gamma$ grain size below which $\varepsilon$ detections are unlikely.
The detection limit of $\varepsilon$ is set as the minimum recorded $\varepsilon$ grain diameter of \qty{\sim 5}{\micro\metre}.
Next, the mean ratio of ($\varepsilon$ child):($\gamma$ parent) diameters was determined to be \num{\sim 0.4}.
The minimum $\gamma$ grain diameter above which $\varepsilon$ nucleation could theoretically be detected is therefore equal to \qty{\sim 12.5}{\micro\metre}.
By taking the cumulative distribution function of all recorded $\gamma$ grain diameters, it was found that \qty{\sim 98}{\percent} of all $\gamma$ grains were larger than this critical diameter.
Therefore, it is determined that the detection limits are not the primary cause of the difference in grain size distributions, and the difference is attributed to reduced transformation stability in larger grains.

\subsubsection{Grain neighbourhood}
\label{sec:dis:neighbourhood}
Given the influence of the $\alpha'$ martensite transformation on the stress state of neighbouring ferrite grains previously identified in the EBSD data, it is prudent to explore the inverse -- whether the neighbourhood configuration can influence the stability of $\gamma$ grains.
Figure~\ref{fig:with_without_dists}b contains the results of this exploration in the 3DXRD dataset - grains with a more ferrite-dense neighbourhood seemed to be more likely to contain detectable $\varepsilon$ at a given load step.
This finding is supported by prior explorations of the EBSD data, with pre-existing athermal $\alpha'$ increasing the driving force for transformation and therefore reducing $\gamma$ grain stability.
This finding has some established literature support -- variability in austenite grain stability has been cautiously assigned to strain localisation in the immediate neighbourhood \citep{toda_multimodal_2022}.
\citet{el_hachi_multi-scale_2022} recently found, using 3DXRD, that martensitic transformation in unfavourably-oriented grains was promoted by transformation in neighbouring grains.
In-situ XRD studies of $\alpha'$ initiation by \citet{hedstrom_stepwise_2007} have also proposed that existing $\alpha'$ can promote direct autocatalyic ($\gamma \rightarrow \alpha'$) transformation in neighbouring grains -- this study demonstrates its role in the two-step ($\gamma \rightarrow \varepsilon \rightarrow \alpha'$) transformation, suggesting that the governing $\alpha'$ nucleation mechanism is stress localisation.

Figure~\ref{fig:with_without_dists}a has shown a relationship between grain size and $\gamma$ grain stability.
Figure~\ref{fig:100_angle_vs_vm}b also explores the relationship between $\gamma$ grain stability against the relative size of the $\gamma$ grain compared with the mean grain size of its immediate neighbours.
A clear inverse relationship is seen ($p < 0.001$), where $\gamma$ grains significantly smaller than their neighbours are more stable against transformation.
This supports prior observations by \citet{tirumalasetty_deformation-induced_2012}, who found that austenite grains positioned near multiple ferrite grains tended to transform earlier, but austenite grains entirely embedded inside larger ferrite grains were more stable against transformation.

\subsection{Martensite variant prediction}
\begin{figure}
    \centering
    \includegraphics[width=\columnwidth]{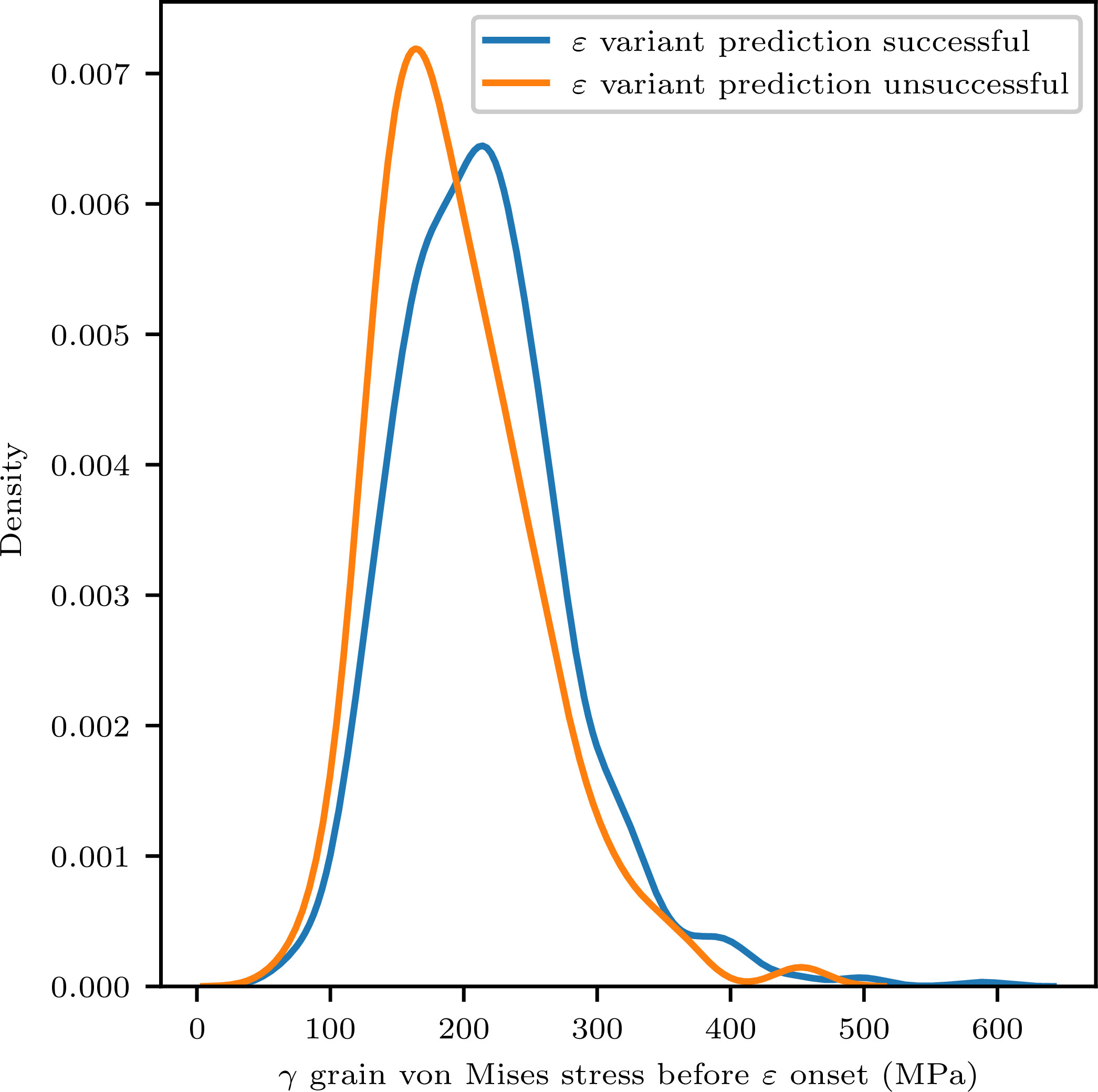}
    \caption{Comparison of $\gamma$ grain von Mises stress in $\gamma$ grains where the $\varepsilon$ variant was successfully vs unsuccessfully predicted.}
    \label{fig:e_variant_prediction_vs_vm_stress_before_formation}
\end{figure}

For the first time, the $\varepsilon$ martensite variant has been both predicted and detected for a large number of $\varepsilon$-forming $\gamma$ grains under in-situ deformation.
The ability to non-destructively identify a large number of per-grain crystallographic orientations is largely limited to the far-field 3DXRD technique employed in this paper.
The Shōji-Nishiyama variant of each $\varepsilon$ grain was both determined and predicted from the orientation of the parent $\gamma$ for each tracked $\gamma$ grain, using the macroscopic uniaxial ($zz$) stress tensor.
The variant prediction routine was able to accurately predict the $\varepsilon$ variant given the parent $\gamma$ orientation for \qty{71.7}{\percent} of tracked $\gamma$ grains.
To determine the factors that influenced whether a variant prediction would be successful, the tracked $\gamma$ grains list was split into two groups based on whether or not the variant prediction succeeded.
Then, the histograms of the grain-level von~Mises stress were plotted for each group in Figure~\ref{fig:e_variant_prediction_vs_vm_stress_before_formation}.

There are a large number of variant prediction models that have been employed, primarily to the direct $\gamma \rightarrow \varepsilon$ transformation.
The reader is referred to a recent study by \citet{tomida_variant_2018} that collated and evaluated a number of these selection mechanisms.
In the scope of the current work, the "minimum strain work" criterion proposed by \citet{humbert_modelling_2002} has been evaluated.
This model has been experimentally verified in small numbers of EBSD grains \citep{lee_variant_2005, wang_ebsd_2021} and grains captured with XRD micro-diffraction \citep{li_situ_2014}.
In this study, the model has been probed on a much larger scale, and a high proportion (\qty{71.7}{\percent}) of $\varepsilon$ formation events were shown to agree with model predictions, demonstrating a high level of model performance.
More recent studies by \citet{tomida_variant_2018, yamashita_martensitic_2018, wang_ebsd_2021} have found deviations to the minimum strain work model predictions, replicating the findings of this study. In this prior literature, deviations were attributed to complexities in the local stress field, as well as favourable minimisation of the interfacial energy to neighbouring grains.
The findings of the current study, shown in Figure~\ref{fig:e_variant_prediction_vs_vm_stress_before_formation}, support these assumptions --  $\gamma$ grains with a successful $\varepsilon$ variant prediction tended to have a greater grain-level $\varepsilon$-onset von~Mises stress than $\gamma$ grains without, indicating the local neighbourhood effect can modify grain stresses to the extent where variant predictions that rely on the global stress state will fail.
These local neighbourhood effects are likely governed by Type III (intragranular) stresses, that alter the local stress field sufficiently to determine which variant forms.
This type of stress concentration is evident in regions of high density in the GND maps,  within individual grains, as shown in Figure~\ref{fig:ebsd_interrupted_loading}.
This also explains the lower parent $\gamma$ von~Mises stress seen just before the onset of $\varepsilon$, as shown in Figure~\ref{fig:e_variant_prediction_vs_vm_stress_before_formation} -- these Type III stress concentrations (not observable in the far-field 3DXRD data) promote early $\varepsilon$ in grains which are otherwise less stressed overall, where the grain-averaged Type II stress (observable in the 3DXRD data) is lower.

\section{Conclusions}
A novel complex-phase stainless steel alloy, a derivative of the 304L alloy system, has been developed, with a specified reduction in the retained austenite stability to enable explorations of the deformation-induced martensitic reaction under low applied strains.
Evidence from both in-situ EBSD and 3DXRD have demonstrated this reduced stability, with significant $\gamma \rightarrow \alpha'$ transformations occurring within \qty{10}{\percent} applied strain.
Within the \qty{2}{\percent} strain range observable with far-field 3DXRD, a significant number of $\gamma \rightarrow \varepsilon$ deformation-induced transformation events were captured, for the first time on a per-grain level.
The $\varepsilon$ phase formation followed the Olson-Cohen kinetics model connecting shear band density to applied strain, and had a strong texture.
The $\varepsilon$ phase formation was also found to reduce the axial stress of its parent $\gamma$ grain.
In addition, the grain-level stress evolution of the $\gamma$ and $\alpha$/$\alpha'$ phases was dominated by the applied stresses, but the stress states of the $\varepsilon$ grains were highly variable, indicating that local crystallographic effects dominated the stress response for $\varepsilon$ but not $\gamma$ or $\alpha$/$\alpha'$ phases.
Using $\gamma \rightarrow \varepsilon$ events as a direct measurement of austenite grain stability against transformation, a number of correlations have been observed to microstructural properties:
\begin{enumerate}
    \item \textbf{Grain diameter:} $\gamma$ grains containing $\varepsilon$ at a given load step had much larger diameters than $\gamma$ grains without $\varepsilon$, suggesting that larger $\gamma$ grains are less stable under deformation than smaller $\gamma$ grains.
    \item \textbf{Grain orientation:} $\gamma$ grains with their \hkl{100} directions near-parallel to the load direction tended to transform with lower grain-level von~Mises stresses, indicating that they have reduced stability.
    \item \textbf{Grain neighbourhood:} $\gamma$ grains containing $\varepsilon$ had an increased level of BCC/BCT phase ($\alpha$/$\alpha'$) in their immediate neighbourhood on average than $\gamma$ grains without $\varepsilon$. It is evident that a richer "ferritic"/"martensitic" neighbourhood promotes further martensite transformations.
    $\gamma$ grains significantly larger than their immediate neighbours had reduced stability against $\varepsilon$ nucleation.
    In addition, deformation-induced $\alpha'$ formation imposed significant levels of plastic strain to neighbouring polygonal ferrite grains.
\end{enumerate}
The minimum strain work criterion for predicting $\varepsilon$ variants was also evaluated - it was shown to usually accurately predict the $\varepsilon$ variant that eventually formed in a given $\gamma$ grain.
For $\gamma$ grains where the model failed to predict the variant that formed, these exhibited lower grain-level stress states. This is explained by the model using a global stress state, rather than individual grain stresses or the average stress state of the immediate neighbourhood.

\section{Data Availability}
Data supporting this study will be available in 2024 from the ESRF Data Portal \citep{collins_revealing_2024} following an embargo period.

\section{Acknowledgments}
James Ball acknowledges the Diamond Light Source and the University of Birmingham for jointly funding his PhD studentship.
The authors are grateful for the provision and allocation of beamtime at the ESRF, under proposal number MA-4750 \citep{collins_revealing_2024}.
Claire Davis and Carl Slater acknowledge the Engineering and Physical Sciences Research Council for support via grants EP/P020755/01 and EP/V007548/1.

\newpage
\typeout{}

\newpage
\onecolumn
\section*{Supplementary Materials}
\begin{figure*}[ht!]
    \centering
    \includegraphics[width=\textwidth,center]{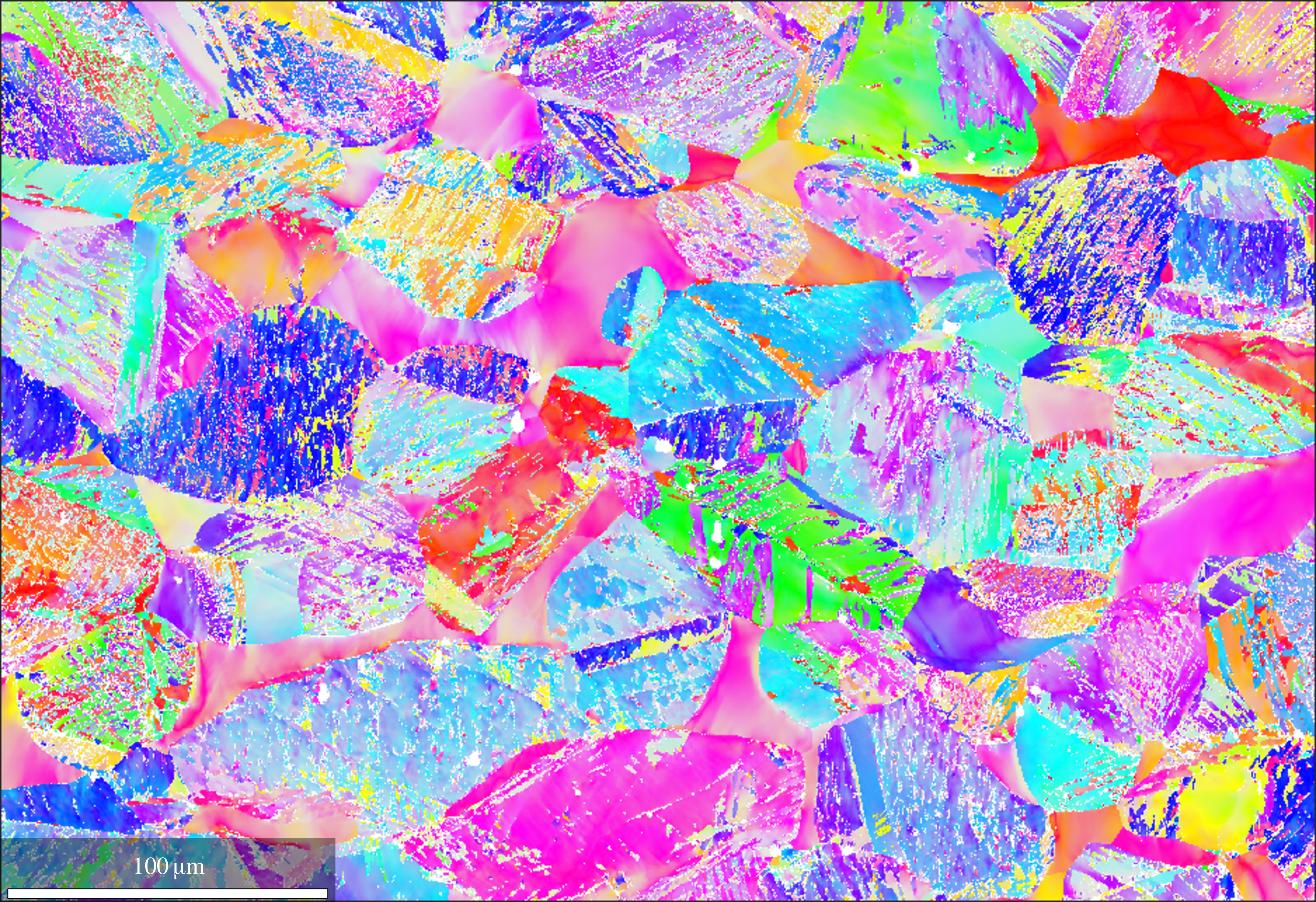}
    \caption{EBSD map of final alloy condition, coloured by IPF-$Z$.}
    \label{fig:load_12_ebsd_ipf}
\end{figure*}

\begin{figure*}[ht!]
    \centering
    \includegraphics[width=\textwidth,center]{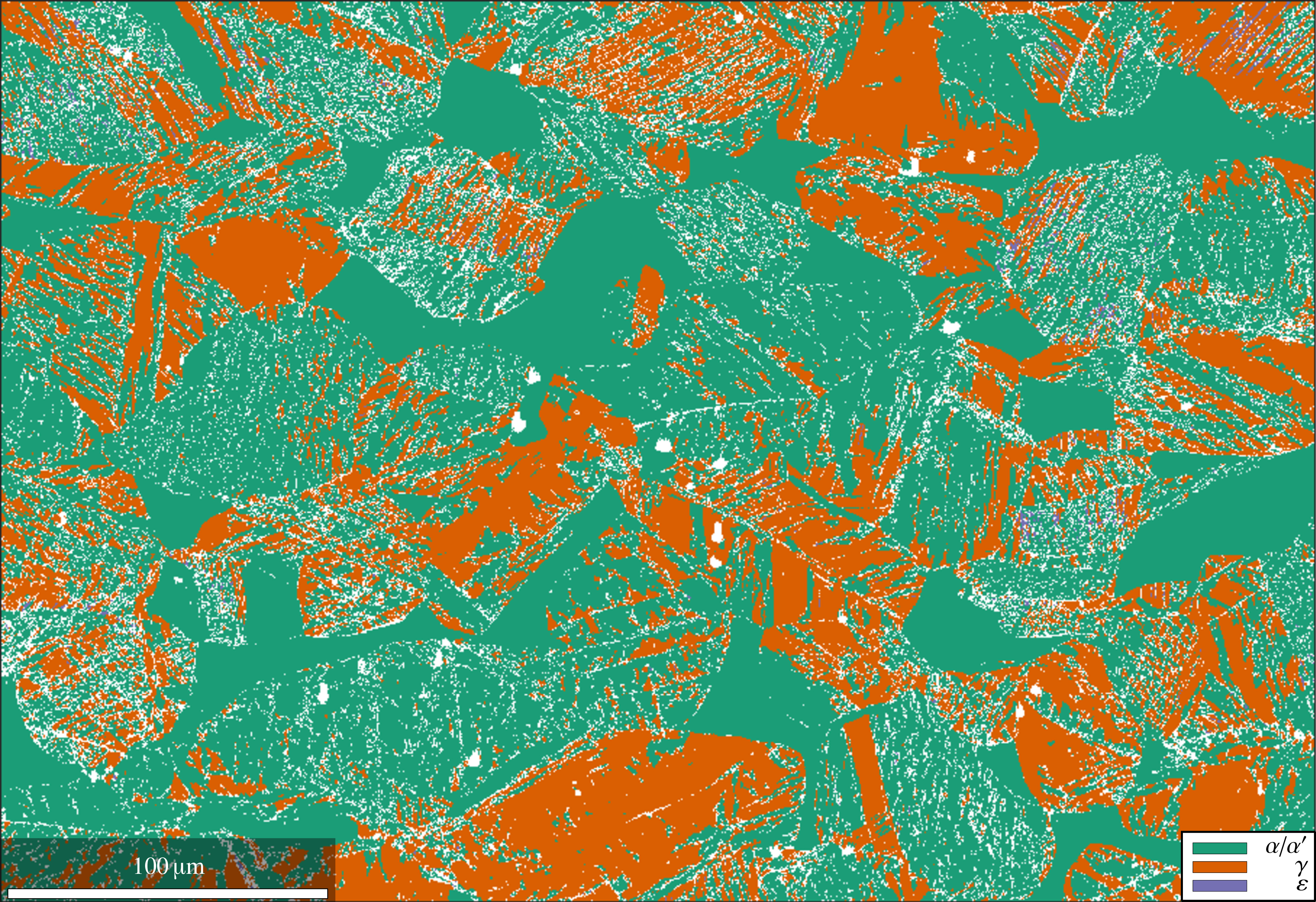}
    \caption{EBSD map of final alloy condition, coloured by phase.}
    \label{fig:load_12_ebsd_phase}
\end{figure*}

\begin{figure*}[ht!]
    \centering
    \includegraphics[width=\textwidth,center]{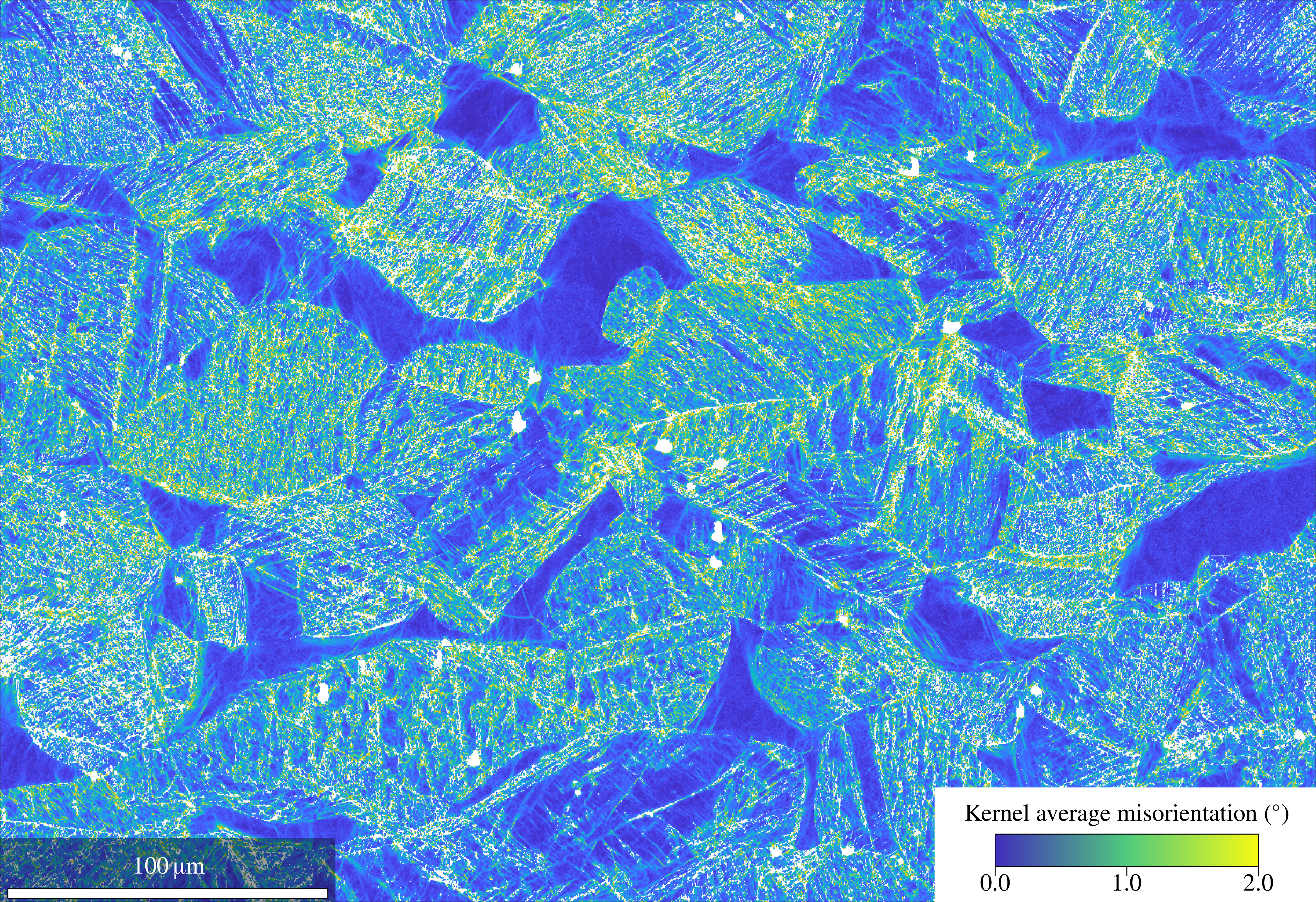}
    \caption{EBSD map of final alloy condition, coloured by kernel-average misorientation.}
    \label{fig:load_12_ebsd_kam}
\end{figure*}

\noindent In this section, the larger-area final EBSD maps of the interrupted-loading EBSD sample are provided, in IPF-Z colouring (Figure~\ref{fig:load_12_ebsd_ipf}), phase colouring (Figure~\ref{fig:load_12_ebsd_phase}) and kernel-average misorientation (KAM) colouring (Figure~\ref{fig:load_12_ebsd_kam}).
A significant amount of plastic deformation is observed in both the IPF and KAM maps, which can be seen most clearly in the polygonal ferrite grains due to the lack of martensite within them.

\end{document}